\documentclass[12pt]{article}
\usepackage{a4}
\usepackage{latexsym}
\usepackage{epsfig}
\begin{document}
\thispagestyle{empty}
\parskip=4mm
\newcommand{\R}{{\mathchoice{\hbox{$\sf\textstyle I\hspace{-.15em}R$}}     
{\hbox{$\sf\textstyle I\hspace{-.15em}R$}}
{\hbox{$\sf\scriptstyle I\hspace{-.10em}R$}}
{\hbox{$\sf\scriptscriptstyle I\hspace{-.11em}R$}}}}
\renewcommand{\theequation}{\arabic{section}.\arabic{equation}}
\renewcommand{\d}{\displaystyle}

\hfill{BUTP-97/14}

\vspace{3cm}
\begin{center}
{\Large\bf Pion Dynamics at Finite Temperature}
\vspace{8mm}

D. Toublan

{\small Institut f\"ur theoretische Physik der Universit\"at Bern\\ CH-3012 
Bern, Switzerland\\

\vspace{3cm}

{\bf Abstract}

\parbox{14cm}{The pion decay constant and mass are computed at low temperature  
within Chiral Perturbation Theory to two loops. The effects of the breaking of 
Lorentz Symmetry by the thermal equilibrium state are discussed. The validity of 
the Gell-Mann Oakes Renner relation at finite temperature is examined.}}
\end{center}

\vfill{
\begin{center}
\rule{36em}{0.02em}
{\footnotesize Work supported in part by Schweizerischer Nationalfonds}
\end{center}}

\setcounter{page}{0}
\newpage
\setcounter{equation}{0}
\section{Introduction}

The spontaneous breaking of chiral symmetry is believed to be a property of 
the strong interaction at zero temperature. If the temperature is finite 
different regimes appear. For sufficiently low temperatures chiral 
symmetry is still spontaneously broken, whereas for high temperatures it has 
to be restored according to asymptotic freedom. The way this 
restoration happens is not yet known and the transition temperature is 
estimated to be around $150-250$ MeV (for a review of QCD at finite temperature 
see~\cite{Smil}).

The low temperature regime hadronic phase is dominated 
by the lightest particle occurring in the spectrum: the pion. The non-zero but 
small masses of the $u$ and $d$ quarks, which explicitly break chiral symmetry, 
make it a pseudo-Goldstone boson of the theory. Because of the lightness of 
these quarks, their masses can be treated as perturbations.

When the system is heated, the first particles to be produced are the pions, 
whereas the states that remain massive in the chiral limit ($m_{\mbox{\tiny 
quarks}} \rightarrow 0$) are exponentially suppressed. The low-energy 
properties of the pions are essentially fixed by chiral symmetry. This 
leads to a wealth of low-energy theorems derived in current algebra at $T=0$. 
For instance the Gell-Mann Oakes Renner (GOR) relation relates the 
pion decay constant and mass to the quark condensate and the quark mass (for 
simplicity  all the quarks are put to the same mass $\hat{m}$)~\cite{GOR}: 
\begin{equation}
\frac{M_{\pi}^2 \; F_{\pi}^2}{\hat{m} \; \langle 0 | \bar{q} q | 0 
\rangle}=-1+O(\hat{m}). \label{eq:go}
\end{equation}
A very efficient technique to analyze the corrections to these theorems is the 
use of effective Lagrangians. The low-energy effective theory of QCD is Chiral 
Perturbation Theory (ChPT)~\cite{Wei,GaL,GaL3}. At zero temperature and non-zero 
quark masses, the GOR relation ceases to hold~\cite{GaL}. The goal of the 
present paper is 
to compute the temperature dependence of the quantities involved in the GOR 
relation and to see whether the latter still holds in the chiral limit.

The thermodynamics of a hadron gas and the quark condensate below the chiral 
phase transition have already been studied up to three loops in 
ChPT~\cite{GeL}. The question of the pion propagation at finite temperature 
has been addressed: the self-energy is known to two loops in 
ChPT~\cite{Sch}, but the pion decay constant to one loop only~\cite{GaLT}. 

The thermal equilibrium state is not Lorentz invariant. In ChPT this first 
explicitly shows up at the two-loop order. It is then important to compute up to 
this order to see what happens in the non-zero temperature case. For instance 
the appearence of two distinct pion decay constants $-$ one for time and one for 
space $-$ begins at this order. This was already noted in~\cite{Pis} and is a 
known feature of some non-relativistic system such as the 
antiferromagnet~\cite{LNR}.

A short presentation of ChPT at finite temperature in the Real Time Formalism  
is the subject of the next section.The two-point axial Green's function is 
computed in Section~3, the pion decay constants and mass in 
Section~4.
The GOR relation at finite temperature is the subject of Section~5. It is 
derived from two Ward Identities relating the axial and pseudoscalar 
two-point Green's functions. Some of the consequences of these identities are 
also examined. The regularization and the chiral limit of T-dependent functions 
appearing in the corpus are appended.

\setcounter{equation}{0}
\section{Cool Chiral Perturbation Theory}

A finite temperature effective theory rests on the zero temperature 
one. In the path integral formalism going from one to another is essentially a 
change of manifold over which one integrates: a torus replaces a 
plane~\cite{TFT}. The Real Time Formalism is used because we are investigating 
time-dependent Green's functions.

To make the presentation simpler, we will only give those results we consider 
necessary here. The reader is invited to consult~\cite{GaL, GaL3} for a 
more detailed discussion on ChPT at $T=0$. We will restrict ourself to the case 
where all quarks 
have the same mass $\hat{m}$.

ChPT is an effective theory describing QCD at low energies. The $N$-flavour 
massless-quark QCD Lagrangian is symmetric under $SU(N)_R\times SU(N)_L$, the 
chiral group. It is assumed that a spontaneous chiral symmetry breakdown 
occurs,
\[SU(N)_R\times SU(N)_L \to SU(N)_V,\]
whose Goldstone bosons are identified as the pions.

The QCD Lagrangian can be approximated at a given order in the momentum using 
an effective Lagrangian expressed in terms of a field $U\in SU(N)$ which 
transforms linearly under $SU(N)_R\times SU(N)_L$,
\[ U\to g_RUg_L^+,\]
and contains the fields of the pseudoscalar Goldstone bosons,
\begin{equation}
U={\rm e}^{{\rm i}\pi^a\tau_a/F},
\end{equation}
where $\tau^a$ are the generators of $SU(N)$ and $F$ is the pion decay 
constant 
in the chiral limit: $F_\pi=F(1+O(\hat{m}))$.

Coupling $U$ with external fields and expanding the effective Lagrangian in 
powers of the external momenta and of quark masses, gives
\begin{equation}
{\cal L}_{\rm eff}={\cal L}^{(2)}+{\cal L}^{(4)}+{\cal L}^{(6)}+\dots 
\label{eq:Leff}
\end{equation}
The promotion of the global chiral symmetry to a local one requires the 
introduction of a derivative $\nabla_\mu U$ which is covariant with respect 
to the external axial and pseudoscalar gauge fields.

To get the desired $O(p^6)$ accuracy, the tree, one- and two-loop diagrams 
of ${\cal L}^{(2)}$, the tree and one-loop graphs of ${\cal L}^{(4)}$ and the 
trees of 
${\cal L}^{(6)}$ are needed.

We will restrict ourselves to the two flavour case ($N=2$) and use the 
tradional notation of~\cite{GaL3}: $\chi=2 B 
\hat{m} \; {\rm I}+2 i B p$, 
$\nabla_{\mu}=\partial_{\mu} U-i \{a_{\mu},U\}$ and 
$F^{\mu\nu}_{R,L}=\pm \partial^{\mu} a^{\nu} \mp \partial^{\nu} a^{\mu}-i 
[a^{\mu},a^{\nu} ]$, where $B$ is proportional to the quark condensate in the 
massless quark limit. It is found that the parts of the Lagrangian 
that contribute to the axial and pseudoscalar two-point Green's functions to 
$O(p^6)$ are: 

1) Lowest order:
\begin{equation}
{\cal L}^{(2)}=\frac{F^2}{4} \langle \nabla_{\mu} U^{\dagger}  \nabla^{\mu} 
U +\chi U^{\dagger} +\chi^{\dagger} U \rangle \label{eq:L2}
\end{equation} 

2) Second order:
\begin{eqnarray}
{\cal L}^{(4)} & = & L_1 \langle \nabla_{\mu} U^{\dagger}  \nabla^{\mu} U 
\rangle^2+
L_2 \langle \nabla_{\mu} U^{\dagger}  \nabla_{\nu} U \rangle \langle 
\nabla^{\mu} U^{\dagger}  \nabla^{\nu} U \rangle \nonumber \\
&&+L_4 \langle \chi U^{\dagger} 
+\chi^{\dagger} U \rangle \langle \nabla_{\mu} U^{\dagger}  \nabla^{\mu} U 
\rangle + L_6 \langle \chi U^{\dagger} +\chi^{\dagger} U \rangle ^2  \\ 
&&+L_8 \langle \chi U^{\dagger} \chi U^{\dagger} +\chi^{\dagger} U 
\chi^{\dagger} 
U \rangle+ L_{10} \langle U^{\dagger} F_R^{\mu \nu} U F_{L \mu \nu} \rangle 
\nonumber \\
&&+ H_1 \langle  F_R^{\mu \nu}  F_{R \mu \nu} +F_L^{\mu \nu}  F_{L \mu 
\nu}\rangle  
+H_2 \langle \chi^{\dagger} \chi \rangle \nonumber \label{eq:L4}
\end{eqnarray} 

3) Third order: the singular part of the two-point function we are interested in 
only receive $T$-independent corrections from the trees of ${\cal L}_{\rm eff}$ 
(see Section~3 for an illustration of this property at one loop). Hence ${\cal 
L}^{(6)}$ won't be explicitly needed here.

When the temperature $\beta^{-1}$ is non-zero, the fields $U(x)$ map the 
$\beta$-dependent torus into $SU(N)$. The generating functional of the 
connected 
Green's functions is:
\begin{equation}
\exp \Big\{i Z[a_{\mu}, p] \Big\}=\int [{\rm d} U] \exp \left\{ i \int_C d^3x \; 
d \tau {\cal L}_{eff} \right\}. \label{eq:Z}
\end{equation}
The integration extends over $\R^3$ and a contour $C$. In the Real Time 
Formalism of quantum field theory at finite 
temperature, which has been thoroughly studied in~\cite{TFT}, one can choose 
different integration paths in the complex 
$t$-plane. We take the so-called Keldysh path shown in Figure 1. The 
functional integral extends over field configurations with the boundary 
condition $U(-\infty,\vec{x})=U(-\infty-i \beta,\vec{x})$, which is the 
familiar 
periodicity condition. A generalization of time-ordering has to 
be introduced: $T$ is replaced by $T_C$, an operator which orders operators 
according to the occurence of their time-arguments on the contour. A Heaviside, 
$\theta_C$, and a Dirac, $\delta_C$, distributions are defined on the 
contour~\cite{TFT}.

The thermal propagator is a Green's function on the contour:
\begin{equation}
(\Box_C+M^2) \; D_{\beta} (x-y) = \delta_C (x-y).
\end{equation}
The unique solution satisfying the Kubo-Martin-Schwinger boundary condition 
is
\begin{eqnarray}
D_{\beta} \left( \tau-\tau'; \omega_k \right)&=&-\frac{i}{2 
\omega_k} \; e^{\beta \omega_k} \; n_B(\omega_k)  \nonumber \\ 
&& \hspace{-1.5cm} \left\{  \left[  e^{-i \omega_k (\tau-\tau')}+e^{-\beta 
\omega_k+i \omega_k 
(\tau-\tau')} \right] \theta_C (\tau-\tau') \right. \\
&&\left. \hspace{-1.3cm} + \left[ e^{i \omega_k (\tau-\tau')}+e^{-\beta 
\omega_k-i \omega_k 
(\tau-\tau')} \right] \theta_C (\tau'-\tau) \right\}, \nonumber
\end{eqnarray}
where $n_B(x)=1/(e^{\beta x}-1)$ is the Bose distribution and 
$\omega_k=\sqrt{\vec{k}^{\; 2}+M^2}$. 

When~(\ref{eq:Z}) is used to compute Green's functions with arguments on the 
real $t$-axis, only a part of $Z[a_\mu,p]$ is relevant: the $C_1$ and $C_2$ 
contour 
segments are the important ones.

The introduction of the usual $2 \times 2$ formalism allows us to rewrite the 
generating functional as ($j=a_\mu,p$):
\begin{equation}
Z[j_1,j_2]=\int [{\rm d} U_1] [{\rm d} U_2] \; \exp \left\{ i \int \frac{1}{2} 
\; 
U_m D_{\beta}^{-1 \; m n} U_n -V \left[U_1 \right]+V \left[ U_2 \right] + j_n 
U_n \right\}, \label{eq:Z12}
\end{equation}
where $m,n=1,2$, and $j_1(x)=j(x)$, $j_2(x)$ are independent sources.

The real-time Green's functions are generated by differentiating~(\ref{eq:Z12}) 
with 
respect to $j_1$ and setting both $j_1$ and $j_2$ to zero. For 
instance the real time two-point axial Green's function is given by:
\begin{eqnarray}
i \langle T A^a_{\mu}(x) A^b_{\nu}(y) \rangle_T &=& \frac{1}{ {\rm Tr} 
e^{-\beta 
H}} \; {\rm Tr} \left\{ e^{-\beta H} T A^a_{\mu}(x) A^b_{\nu}(y) \right\} 
\nonumber \\
&=&\left. \frac{\delta^2}{\delta a_{1 \mu}^a(x) \; \delta a_{1 \nu}^b(y)} \; 
Z[a_{1 \mu},a_{2 \mu},p_1,p_2] \right|_{a_{1 \mu}=a_{2 \mu}=p_1=p_2=0}.
\end{eqnarray}

In momentum space the propagator $D^{m n}_{\beta}$ reads:
\begin{equation} 
\left\{ 
      \begin{array}{l}
      i D^{11}_{\beta}(k)=\left[ i D^{22}_{\beta}(k)                             
\right]^*=\tilde{\Delta}(k)+
        2 \pi \; \delta(k^2-M^2) \; n_B (|k_0|)=:\tilde{\Delta}_{\beta}(k) \\ 
       i D^{12}_{\beta}(k)= i D^{21}_{\beta}(k) \; e^{-\beta \; |k_0|}=2 \pi 
       \; \delta(k^2-M^2) \; n_B (|k_0|)
      \end{array}
\right., \label{eq:prop}
\end{equation}
where $\tilde{\Delta}(k)=i/(k^2-M^2+i \varepsilon)$ is the usual Feynman 
propagator.

The topological and combinatorical structures of the Feynman diagrams do 
not change, but the propagator has now acquired a matrix structure and there 
are two kinds of fields. In a real-time Green's function only the type $1$ 
field
can appear on an external leg and the type $2$ field plays the role of a ghost 
field. There are also two kinds of vertices, which are equivalent up to a sign. 
The two different fields interact through the off-diagonal terms of the 
propagator.

\setcounter{equation}{0}
\section{The axial two-point Green's function}

The finite temperature two-point Green's function of the axial current
\begin{equation}
A^a_{\mu}(x)=\bar{q}(x) \frac{\tau^a}{2} \gamma_{\mu} \gamma_5 q(x) 
\label{eq:ax}
\end{equation}
describes the dynamics of a pion 
in a strongly interacting gas ($\tau^a$ are the $SU(2)$ generators). Let $G_{\mu 
\nu}(x-y,T) \; \delta^{a b} :=i  \langle 
T \; A^a_{\mu}(y) A^b_{\nu}(y) \rangle_T$. In general the two-point axial 
Green's function at finite temperature can be written as:
\begin{equation}
\tilde{G}_{\mu \nu}(q,T)=-\frac{q_{\mu} q_{\nu} \alpha^A(q,T)+q_{\mu} 
\beta^A_{\nu} (q,T)+q_{\nu} \beta^A_{\mu}(q,T)+\gamma^A_{\mu 
\nu}(q,T)}{q_0^2-\Xi(q,T)}+\rho^A_{\mu \nu}(q,T) \label{eq:GA}
\end{equation}
It contains a singular piece and a finite part, $\rho_{\mu\nu}^A(q,T)$. To get 
this Green's function to order $T^4$ in $SU(2)$-ChPT (two 
flavours), one needs to compute the Feynman graphs shown in Figure 2 with the 
effective Lagrangian (\ref{eq:Leff}).

The singular parts of the tree diagrams are temperature independent. They just 
contain the zero temperature Feynman propagator $\tilde{\Delta}(q)$.  Most of 
them have already been computed in~\cite{GaL,Bur}. The other components of the 
matrix propagator generate imaginary parts of the finite term of the two-point 
Green's function. Thus they are not interesting in our context. They will only 
be mentioned in the one-loop computation given below and ignored elsewhere.

The $T$-dependence of the one-loop graphs is rather simple: only properties of 
the thermal propagator at the origin appear in that case~\cite{GaL}.

At lowest order the pion mass is
\begin{equation}
M^2:=2 \hat{m} B 
\end{equation}
and the various $O(p^4)$ graphs are: 
\begin{eqnarray}
\tilde{G}_{\mu \nu}^{(4.1+4)}(q,T)&=&g_{\mu \nu} \left(F^2 +16 M^2 L_4-2 
\Delta_{\beta}(0)  \right) \\
&&  \hspace{1.5cm}+[q^2 \; g_{\mu \nu}-q_{\mu} q_{\nu}] \; \left( 4 H_1-2 
L_{10} 
\right)  \nonumber \\
\tilde{G}_{\mu \nu}^{(4.2+5)}(q,T)&=&q_{\mu} q_{ \nu} \left(\frac{8}{3} 
\Delta_{\beta}(0)+
32 L_4 M^2 \right) \tilde{\Delta}_\beta(q)\\
\tilde{G}_{\mu \nu}^{(4.3+6)}(q,T)&=&i \; q_{\mu} q_{ \nu} \Bigg(\frac{4 
q^2-M^2}{6} \; \Delta_{\beta}(0)  \label{eq:sqdelta} \\
&&  \hspace{.2cm} +M^2 (-16 L_4 q^2+32 L_6 M^2+16 L_8 M^2) \Bigg) 
\Big[(\tilde{\Delta}_\beta(q))^2-(iD_\beta^{12}(q))^2 \Big]. \nonumber
\end{eqnarray}
The relation $( \Box +M^2) \Delta_\beta(x)=i \delta^{(d)}(x)$ has been used to 
simplify the expressions.

At first sight, the term in square brackets in (\ref{eq:sqdelta}) seems to be 
ill defined: it contains products of Dirac distributions with the same argument. 
But the equation of motion implies that
\begin{eqnarray}
\partial_{M^2} \Delta_{\beta}(x)&=&i \frac{d}{d M^2} \; D_{\beta}^{11}(x)  
\label{eq:dmdbx} \\
&&\hspace{-1cm} =-i \int d^dy \; \left( 
D_{\beta}^{11}(x-y) \; D_{\beta}^{11}(y)- D_{\beta}^{12}(x-y) \; 
D_{\beta}^{12}(y)\right). \nonumber 
\end{eqnarray}
Hence the products of thermal statistical weights at intermediate stages of the 
computation disappear when the different types of vertices are combined. This 
is a generic feature of the Real Time Formalism~\cite{TFT}.

Moreover the terms coming from the thermal parts of $\Delta_\beta(q)$ and 
the alike only contribute to $\rho_{\mu\nu}^A(q,T)$, the finite term of the 
two-point function. 
To get the pole and residue of the latter, it suffice to replace 
$\Delta_\beta(q)$ by the corresponding Feynman propagator everywhere. This is 
the case to all orders in the perturbation.

The functions appearing 
in~(\ref{eq:GA}) 
at first non-leading order are already known~\cite{GaLT}:
\begin{eqnarray}
\Xi(q,T)&=&\vec{q}^{\; 2}+M^2 \left\{ 1+\frac{M^2}{F^2} \left[ -16 L_4+32
L_6+16 L_8 \right] \right\}  \\
&& \hspace{1cm} +\frac{M^2}{2 F^2} \Delta_{\beta}(0) +O(p^6),
\nonumber \\
\alpha ^A (q,T)&=&F^2 \left\{ 1+\frac{M^2}{F^2} \left[ 16 L_4 \right] \right\}
-2 \Delta_{\beta}(0)+O(p^6),\\
\beta^A_{\mu}&=&O(p^6), \hspace{4cm} 
\gamma^A_{\mu \nu}=O(p^6), \nonumber \\
\rho^A_{\mu \nu}(q,T)&=&g_{\mu \nu} \left(F^2 +16 M^2 L_4-2 
\Delta_{\beta}(0) \right) \\
&&  \hspace{1.5cm}+[q^2 \; g_{\mu \nu}-q_{\mu} q_{\nu}] \; \left( 4 H_1-2 
L_{10} \right) +O(p^6).  \nonumber
\end{eqnarray}

Only $\rho^A_{\mu\nu}(q,T)$ really depends on the momentum. As 
already said, the Lorentz symmetry is not explicitly broken at this order.
 
Because $\Delta_{\beta}(0)$ contains the Feynman propagator at the origin,
it diverges. As usual in ChPT, because it preserves the symmetry of 
the theory, dimensional regularization is used~\cite{GaL,GeL}:
\begin{equation}
\Delta_{\beta}(0)=\int \frac{d^d k}{(2 \pi)^d} \tilde{\Delta}_{\beta}(k)
=2 M^2 \lambda+{\cal N}_M(T), \label{eq:db}
\end{equation}
where \begin{equation}
\lambda=\frac{\mu^{d-4}}{(4 \pi)^2} \left( \frac{1}{d-4} 
-\frac{1}{2}[\ln 4 \pi+\Gamma^{\prime}(1)+1] + \ln (\frac{M}{\mu}) 
+O(d-4) \right), \label{eq:lamb}
\end{equation}
and \begin{equation}
{\cal N}_M(T)=\int \frac{d^4 k}{(2 \pi)^3} \delta (k^2-M^2) \; n_B(\omega_k).
\label{eq:nm}
\end{equation}

The renormalization of the theory at finite temperature has to be the same as
at $T=0$. This is obviously here the case: no divergence occur in the 
temperature dependent part of the two-point function.
The scale independent parameters used to renormalize the theory are: 
\begin{equation}
\bar{L}_i=L_i-\gamma_i \lambda, \hspace{3cm} \bar{H}_i=H_i-\delta_i \lambda,
\end{equation}
where $\gamma_1=1/12$, $\gamma_2=1/6$, $\gamma_4=1/4$, $\gamma_6=3/32$, 
$\gamma_8=0$, $\gamma_{10}=2/3$, $\delta_1=1/3$ and $\delta_2=0$. In ChPT at 
$T=0$, the expansion of the pion mass and decay 
constant at $O(p^4)$ can be expressed in terms of  $F$, $M$ and $\bar{L}_i$:
\begin{eqnarray}
M^2_{\pi}&=& M^2 \left \{1+ \frac{M^2}{F^2} \left[ -16 
\bar{L}_4+32\bar{L}_6+16 \bar{L}_8 \right] \right\}+O(p^6),\\
F^2_{\pi}&=&F^2 \left \{1+ 
\frac{M^2}{F^2} \left[ 16 \bar{L}_4\right] \right\}+O(p^6) \label{eq:MF}
\end{eqnarray}

The $O(p^6)$ graphs are a bit more complicated than those of $O(p^4)$.
However, only the $T$-dependent terms are of interest here. Thus the diagrams
$(6.23-28)$ will not be explicitly given (see~\cite{Bur,L6,Bij} for an analysis 
of zero temperature ChPT to $O(p^6)$).

Some of the two-loop graphs, $(6.1-3,6.13-16)$, are just
products of lower order ones.

The diagrams $(6.4-6,6.17-19)$ involve essentially the same elements as those 
occuring at one loop. Only their vertices are different: they contain some 
low-energy coupling constants and derivatives. For instance the graphs $(6.5)$ 
and $(6.18)$ give:
\begin{eqnarray}
\tilde{G}_{\mu \nu}^{(6.5)}&=&-\frac{8 i}{3 F^2} q_{\mu} q_{\nu} 
\tilde{\Delta}(q) \Delta_{\beta}^2(0), \\
\tilde{G}_{\mu \nu}^{(6.18)}&=&\frac{16 i}{F^2} \tilde{\Delta}(q) \int 
\frac{d^dk}{(2 \pi)^d} \; \tilde{\Delta}_{\beta}(q) \; \Big\{ (2 L_1+4 L_2) \; 
kq \; (k_{\nu} q_{\mu}+k_{\mu} q_{\nu})   \\
&& \hspace{2cm} + (6 k^2 L_1+2 k^2 L_2-\frac{17}{3} M^2 L_4)  \; q_{\mu} 
q_{\nu} 
 \Big\} \nonumber.
\end{eqnarray}

The graphs $(6.7-9,6.20-22)$ are special ones for the Real Time Formalism. 
They are the only diagrams in the whole set which contribute to the pole and 
residue of the two-point function and contain type $2$ fields. Their role 
is very important for the consistency of the
theory~\cite{TFT}. The graphs $(6.8)$ and $(6.21)$ are taken as examples:
\begin{eqnarray}
\tilde{G}_{\mu \nu}^{(6.8)}&=&-\frac{4}{9 F^2} \; q_{\mu} q_{\nu} \; 
\tilde{\Delta}(q)  \; \Delta_{\beta}(0) \nonumber \\
&& \hspace{1.5cm} \int \frac{d^dk}{(2 \pi)^d} (4 k^2-M^2) \left[ \left(i 
D_{\beta}^{11}(k) \right)^2- \left(i D_{\beta}^{12}(k) \right)^2 \right] 
\label{eq:6.8},
\end{eqnarray}
where the fact that the type-2 tadpole is the same as the type-1 tadpole has 
been used, and
\begin{eqnarray}
\tilde{G}_{\mu \nu}^{(6.21)}&=&\frac{128 M^2}{3 F^2} \; q_{\mu} q_{\nu} \; 
\tilde{\Delta}(q) \; \int \frac{d^dk}{(2 \pi)^d} \; (k^2 L_4-2 M^2 L_6-M^2 L_8) 
\nonumber \\
&& \hspace{5cm}  \left[ \left(i D_{\beta}^{11}(k) \right)^2- \left(i 
D_{\beta}^{12}(k) \right)^2 \right]. \label{eq:6.21}
\end{eqnarray}

As a consequence of~(\ref{eq:dmdbx}), the integrals in 
(\ref{eq:6.8},\ref{eq:6.21}) are essentially 
$\partial_{M^2} \Delta_{\beta}(0)$. In $d=4$, this quantity contains a singular 
piece:
\begin{equation}
\partial_{M^2} \Delta_{\beta}(0) = 2 \lambda+\frac{1}{16 \pi^2}+\partial_{M^2} 
{\cal N}_M(T).
\end{equation}

Finally, we turn to the genuine two-loop graphs. These
are the closed-eye $(6.10)$, the guimbard $(6.11)$ and the sunset $(6.12)$, 
given by:
\begin{eqnarray}
\tilde{G}_{\mu \nu}^{(6.10)}(q,T)&=&-\frac{4}{9 F^2} \int \frac{d^dk_1 \; 
d^dk_2}{(2 
\pi)^{2d}} \; \tilde{\Delta}_{\beta} (k_1) \; \tilde{\Delta}_{\beta} (k_2) \; 
\tilde{\Delta}_{\beta} (q-k_1-k_2) \nonumber \\
&& \hspace{1.5cm} \left\{ (3 k_{1 \mu} +3 k_{2 \mu}-2 q_{\mu}) \; (3 k_{1 \nu} 
+3 k_{2 \nu}-2 q_{\nu}) \right\}, \label{eq:ces}
\\
\tilde{G}_{\mu \nu}^{(6.11)}(q,T)&=&\frac{2}{9 F^2} \;  q_{\mu} 
\tilde{\Delta}(q) \int 
\frac{d^dk_1 \; d^dk_2}{(2 
\pi)^{2d}} \; \tilde{\Delta}_{\beta} (k_1) \; \tilde{\Delta}_{\beta} (k_2) \; 
\tilde{\Delta}_{\beta} (q-k_1-k_2)  \nonumber \\
&& \hspace{-1cm} \left( 2 q_{\nu} -3 k_{1 \nu}-3 k_{2 \nu} \right) \left\{ 
k_1^2+k_2^2+4 k_1 
k_2+M^2+2 (k_1+k_2)q-2q^2 \right\}  \nonumber \\
&&+ \Big( \; \mu \; \longleftrightarrow \; \nu \; \Big), \label{eq:guis} \\
\tilde{G}_{\mu \nu}^{(6.12)}(q,T)&=&-\frac{i}{18 F^2} \; q_{\mu} q_{\nu} 
\tilde{\Delta}^2(q) \int 
\frac{d^dk_1 \; 
d^dk_2}{(2 \pi)^{2d}} \; \tilde{\Delta}_{\beta} (k_1) \; \tilde{\Delta}_{\beta} 
(k_2) \; 
\tilde{\Delta}_{\beta} (q-k_1-k_2) \nonumber \\
&& \hspace{.3cm} \left\{ 3 M^4+ \left(  k_1^2+k_2^2+4 k_1 k_2+M^2+2 
(k_1+k_2)q-2q^2 \right)^2 \right\}. \label{eq:suns}
\end{eqnarray}

One readily sees that the polynomials appearing in the integrands are 
essentially four-point functions at tree level. For instance the 
square of the tree-level isospin averaged  $\pi$-$\pi$ scattering amplitude 
appears in the sunset. These integrals may first appear complicated, but using 
the symmetry properties of the integrands they can be expressed in terms of two 
independent integrals (cf Appendix~A for more details):
\begin{equation}
\tilde{G}_{\mu \nu}^{(6.10)}(q,T)= -\frac{4}{9 F^2} \; q_{\mu} q_{\nu} \; 
I(q,T) 
+ \frac{9}{F^2} \; I_{\mu \nu}(q,T), 
\end{equation}
\begin{eqnarray}
\tilde{G}_{\mu \nu}^{(6.11)}(q,T)&=& \frac{\tilde{\Delta}(q)}{F^2} \left( 
\frac{8}{9} q_{\mu} q_{\nu} q^2 I(q,T)-4 q_{\mu} q^{\rho} I_{\nu \rho}(q,T) 
\right. \\
&& \left. \hspace{2cm}-4 q_{\nu} q^{\rho} I_{\mu \rho}(q,T)+ 4 q_{\mu} q_{\nu} 
\Delta_{\beta}^2(0) \right), \nonumber 
\end{eqnarray}
\begin{eqnarray}
\tilde{G}_{\mu \nu}^{(6.12)}(q,T)&=& -q_{\mu} q_{\nu} \; 
\frac{\tilde{\Delta}^2(q)}{18 F^2} \left\{ (5 M^4-8 q^4) 
I(q,T) \right. \\
&&\left. \hspace{2cm} +72 q^{\rho} q^{\sigma} I_{\rho \sigma}(q,T)-12 (M^2-3 
q^2) \Delta_{\beta}^2(0) \right\}. \nonumber
\end{eqnarray}
The functions $I(q,T)$ and $I_{\mu \nu}(q,T)$ are defined as:
\begin{eqnarray}
I(q,T)&=&i \int \frac{d^dk_1 \; 
d^dk_2}{(2 
\pi)^{2d}} \; \tilde{\Delta}_{\beta} (k_1) \; \tilde{\Delta}_{\beta} (k_2) \; 
\tilde{\Delta}_{\beta} (q-k_1-k_2), \label{eq:I} \\
I_{\mu \nu}(q,T)&=&i \int \frac{d^dk_1 \; 
d^dk_2}{(2 
\pi)^{2d}} \; \tilde{\Delta}_{\beta} (k_1) \; \tilde{\Delta}_{\beta} (k_2) \; 
\tilde{\Delta}_{\beta} (q-k_1-k_2) \; k_{1 \mu} \; k_{1 \nu}. \label{eq:Imn}
\end{eqnarray}
These integrals have to be regularized because they are divergent in $d=4$. 
Their finite parts are determined by the four functions ${\cal N}_M (T)$, 
$N_{\mu \nu}(M,T)$, $\bar{I}(q,T)$ and $\bar{I}_{\mu \nu}(q,T)$, given in 
Appendix~A. 

The end result for the various terms appearing in the 
representation~(\ref{eq:GA}) reads:
\begin{eqnarray}
\Xi (q,T) &=&\vec{q}^{\; 2}+M_{\pi}^2+ \frac{M_{\pi}^2}{2 \; F_{\pi}^2} 
{\cal N}_{M_{\pi}}(T)  \nonumber \\
&& +\frac{1}{F_{\pi}^4} \left[ M_{\pi}^4 \; \bar{L}_{\Xi} \; {\cal 
N}_{M_{\pi}} 
(T) +\frac{M_{\pi}^4}{4} {\cal N}_{M_{\pi}} (T) \; 
\partial_{M_{\pi}^2} {\cal N}_{M_{\pi}} (T)\right. \label{eq:xi} \\
&&  \hspace{1.5cm} \left.     
-\frac{11 M_{\pi}^2}{8} {\cal N}_{M_{\pi}}^2 (T) +\frac{M_{\pi}^4}{6} 
\bar{I}(q,T) - q^{\mu} q^{\nu} \bar{\kappa}_{\mu \nu} (q,T)  \right] +O(p^8), 
 \nonumber
\end{eqnarray}

\begin{eqnarray}
\alpha^A(q,T) &=&F_{\pi}^2-
2{\cal N}_{M_{\pi}}(T)   \label{eq:aa} \\
&& \hspace{-2cm}  +\frac{1}{F_{\pi}^2} \left[  M_{\pi}^2 \; \bar{L}_A \; 
{\cal 
N}_{M_{\pi}}(T)  -M_{\pi}^2 {\cal N}_{M_{\pi}}(T) \; \partial_{M_{\pi}^2} 
{\cal N}_{M_{\pi}}(T) +2 {\cal N}_{M_{\pi}}^2(T)  
\right]  +O(p^8), \nonumber 
\end{eqnarray}

\begin{equation}
\beta^A_{\mu}(q,T) =\frac{1}{F_{\pi}^2} \; q^{\nu} \bar{\kappa}_{\mu \nu} 
(q,T)+O(p^8), 
\label{eq:ba} 
\end{equation}

\begin{equation}
\gamma^A_{\mu \nu}=O(p^{10}), \label{eq:ga}
\end{equation}

\begin{eqnarray}
\rho^A_{\mu \nu}(q,T)&=&R^A_{\mu \nu} (q)+ g_{\mu \nu} \; \alpha^A(q,T) 
\nonumber \\
&&\hspace{-1.5cm} +\frac{1}{F_{\pi}^2} \; \left[ \bar{\kappa}_{\mu \nu}(q,T)+ 
\left( 
q_{\mu} q_{\nu} - g_{\mu \nu} \; q^2 \right) \; \left( 2 
\bar{L}_{10}-\frac{1}{72 \pi^2} \right) \right]+O(p^8).
\label{eq:ra}
\end{eqnarray}
The function
\begin{equation}
\bar{\kappa}_{\mu \nu} (q,T):=\bar{L} \; N_{\mu \nu} (M_{\pi},T)  +4  
\bar{I}_{\mu 
\nu} (q,T),
\end{equation}
and different combinations of the renormalized low-energy coupling constants 
were introduced to lighten the expressions:
\begin{equation}
\left\{
\begin{array}{lll}
\bar{L}&=&32 \bar{L}_1+ 64 \bar{L}_2-\frac{7}{18 \pi^2} \\
\bar{L}_{\Xi}&=&-48 \bar{L}_1-16 \bar{L}_2 +48 \bar{L}_4-80 \bar{L}_6 -40 
\bar{L}_8+\frac{55}{576 \pi^2} \\
\bar{L}_A&=& 48 \bar{L}_1+16 \bar{L}_2-24 \bar{L}_4-\frac{7}{144 \pi^2}. \\
\end{array} \right. \label{eq:lambda}
\end{equation}
Expressed in terms of the coupling constants defined in~\cite{GaL}, these are 
given by
\begin{equation}
\left\{
\begin{array}{lll}
\bar{L}&=&(\bar{l}_1+4\bar{l}_2-\frac{14}{3}) / 12 \pi^2  \\
\bar{L}_{\Xi}&=&(-24 \bar{l}_1-16\bar{l}_2+15 \bar{l}_3 + 12 \bar{l}_4 + 
\frac{55}{3}) / 192 \pi^2  \\
\bar{L}_A&=& (6 \bar{l}_1+4\bar{l}_2-9 \bar{l}_4 - \frac{7}{3}) / 48 \pi^2. 
\end{array} \right. 
\end{equation}
Note that $\alpha^A(q,T)$ does not depend on the momentum at this order.

An integral representation can be given for the different functions of 
temperature and momentum appearing in the previous expressions. ${\cal N}_M(T)$ 
has already been introduced in~(\ref{eq:nm}). The other function related to 
the properties of the propagator at the origin is $N_{\mu \nu}(M,T)$. It is 
given in Appendix~A together with the regular parts of the momentum dependent 
functions $I(q,T)$ and $I_{\mu \nu}(q,T)$ defined 
in~(\ref{eq:I},\ref{eq:Imn}). All these functions depend on the ratio $M/T$ in a  
non-trivial way. 

The expressions~(\ref{eq:xi}-\ref{eq:ra}) contain all the 
contributions to the finite temperature axial two-point Green's function to 
$O(p^6)$. Both $M$ and $T$ count as quantities of $O(p)$. The dependence of the 
functions $\Xi(q,T)$, $\alpha^A(q,T)$ and 
$\beta^A_{\mu}(q,T)$ on the external momentum begins at this order. There is an 
important qualitative change between the one-loop and the two-loop results: the 
way the functions involved depend on $q_0$ and on $\vec{q}$ are now different, 
reflecting the breaking of Lorentz symmetry by the heat bath. As it must be, the 
$T\neq0$ renormalization is the same as the one at zero temperature~\cite{GaL}.

\setcounter{equation}{0}
\section{The pion decay constants and mass at finite \boldmath{$T$}}

From the two-point axial Green's function some interesting 
quantities can be derived: the pion decay constants and mass. In the $T=0$ case 
the self-energy is defined as the pole of $\tilde{G}_{\mu \nu}(q,T=0)$ in the 
$q_0$ complex plane and the pion decay constant as its residue at the pole 
position.

The dispersion curve determines the position of the pole in the $q_0$-plane:
\begin{equation}
q_0=\Omega(\vec{q},T).
\end{equation}
In our case a non-trivial momentum dependence occurs at $O(p^6)$. Therefore the 
pole position at the same order can be obtained by replacing $q_0$ by 
$\omega_q=\sqrt{\vec{q}^{\; 2}+M_\pi^2}$, that is:
\begin{equation}
\Omega^2(\vec{q},T)=\Xi(q_0,\vec{q},T) \Big|_{q_0=\omega_q}+O(p^8). 
\label{eq:omega}
\end{equation}

A possible definition of the mass is $M_{\pi}(T):={\rm Re} \; \Omega(\vec{q},T) 
\Big|_{\vec{q}=0}$, i.e. the real part of the pole. 
With the $\bar{L}$s defined in~(\ref{eq:lambda}), the ChPT result reads:
\begin{eqnarray}
M_{\pi}^2(T)&=&M_{\pi}^2 \Bigg\{ \left. 1+\frac{1}{2 F_{\pi}^2} 
{\cal N}_{M_{\pi}} (T)  \right. \nonumber \\
&&\left. +\frac{1}{F_{\pi}^4} \Bigg[ M_{\pi}^2 \left( {\cal N}_{M_{\pi}}(T) \; 
\bar{L}_\Xi +\frac{1}{4} {\cal N}_{M_{\pi}} (T) \; 
\partial_{M_{\pi}^2} {\cal N}_{M_{\pi}} (T) \right.  \right) \nonumber \\
&&  \left. \hspace{1.2cm}+ \frac{M_{\pi}^2}{6} \; {\rm Re} 
[\bar{I}(q,T)] -\frac{11}{8} {\cal N}^2_{M_{\pi}} 
(T) \right.  \label{eq:Mt} \\ 
&& \left. \hspace{1.2cm} -\bar{L} \; N_{00}(M_{\pi},T) -4 \; {\rm 
Re}[ 
\bar{I}_{00}(q,T)] 
\Bigg] \Bigg|_{q_0=M,\vec{q}=0} \right. \Bigg\} +O(p^8). \nonumber 
\end{eqnarray}

This result agrees with~\cite{Sch}, where a somewhat different representation is 
used. The imaginary part of the pole, which determines the damping rate of the 
pions, has been thoroughly studied in the same context: its mean 
approximately behaves like $T^5/F_\pi^5$ above $100$ MeV~\cite{Sch}.

To extract the residue at $O(p^6)$, $\Xi(q,T)$ has to be expanded around the 
pole position:
\begin{equation}
\Xi(q_0,\vec{q},T)=\Omega^2(\vec{q},T)+(q_0-\omega_q) \; \frac{\partial  
\Xi(q,T)}{\partial q_0} \Big|_{q_0=\omega_q}+ \dots
\end{equation}

The thermal equilibrium state is invarian under spatial rotations, it implies 
that
\begin{equation}
\bar{\kappa}_{0i}(q,T)=q_i \; \bar{\kappa}(q,T),
\end{equation}
thus $\beta^A_\mu(q,T)$ can be rewritten as:
\begin{equation}
\beta_0^A(q,T)=q_0 \; \beta_t^A(q,T) \; \; {\rm and} \; \; \beta_i^A(q,T)=q_i \; 
\beta_s^A(q,T). \label{eq:btbs}
\end{equation}
The axial two-point function then takes the form
\begin{equation}
\tilde{G}_{\mu\nu}(q,T)=\Phi^A_{\mu \nu}(q,T)- \frac{f_\mu(q,T) \; 
f_\nu(q,T)}{q_0^2 -\Omega^2(\vec{q},T)}, \label{eq:Gmn} 
\end{equation}
where $\Phi_{\mu \nu}^A(q,T)$ is a finite term and
\begin{equation}
\left\{
\begin{array}{lll}
f_0 (q,T)&=&q_0 \; F_t(\vec{q},T)\\
f_i (q,T)&=&q_i \; F_s(\vec{q},T).
\end{array}
\right.
\end{equation}

The "temporal" residue at the pole position is given by
\begin{equation}
F_t^2(\vec{q},T)=\Big( \alpha^A(q,T)+2 \beta^A_t(q,T) \Big) \; \Big(1+\frac{1}{2 
\omega_q} \; \frac{\partial  \Xi(q,T)}{\partial q_0} \Big) 
\Big|_{q_0=\omega_q}+O(p^8). \label{eq:residue}
\end{equation}
The difference between $\beta_t(q,T)$ and $\beta_s(q,T)$ makes the difference in 
$F_t(\vec{q},T)$ and $F_s(\vec{q},T)$.

The pion decay constants can naturally be defined as $F_{\pi}^{s,t}(T) := 
F_{s,t}(\vec{q},T) \Big|_{\vec{q}=0}$. The ChPT computation gives
\begin{eqnarray}
\left( F_{\pi}^t(T) \right)^2&=&F_{\pi}^2 \left\{ 1-\frac{2}{F_{\pi}^2} 
{\cal N}_{M_{\pi}}(T) \right.  \nonumber \\
&& \left.  +\frac{1}{F_{\pi}^4} \Bigg[ M_{\pi}^2 \left( {\cal 
N}_{M_{\pi}}(T) \; 
\bar{L}_A - {\cal N}_{M_{\pi}} (T) \; 
\partial_{M_{\pi}^2} {\cal N}_{M_{\pi}} (T) \right) 
\right. \nonumber \\
&& \left. \hspace{1cm}   +\frac{M_{\pi}^3}{12} \; 
\frac{\partial}{\partial q_0} \bar{I}(q,T)  +2 {\cal 
N}^2_{M_{\pi}} (T)+\bar{L} \; 
N_{00}(M_{\pi},T)   \right.  \label{eq:Ftt} \\
&& \left.  \hspace{1cm} +4  \bar{I}_{00}(q,T) -2 M_{\pi} 
\frac{\partial}{\partial q_0} 
\bar{I}_{00}(q,T)\Bigg] \Bigg|_{q_0=M,\vec{q}=0} \right\} +O(p^8) \nonumber 
\end{eqnarray}
and the difference between the two pion decay constants is
\begin{eqnarray}
\frac{F_{\pi}^t(T)-F_{\pi}^s(T)}{F_{\pi}}&=&\frac{1}{F_{\pi}^4} \Bigg[ 
\frac{1}{3} \;\bar{L} \; \left( 4 N_{00}(M_{\pi},T) -M_{\pi}^2 {\cal 
N}_{M_{\pi}}(T) \right) +\frac{4}{3} {\cal N}_M^2(T)
   \label{eq:Fst} \\
&& \hspace{-1cm} -\frac{4}{3} M_\pi^2 \bar{I}(q,T)+\frac{16}{3} 
\bar{I}_{00}(q,T)-M_{\pi} \bar{\kappa}(q,T) \Bigg] 
\Bigg|_{q_0=M,\vec{q}=0}+O(p^8). 
\nonumber 
\end{eqnarray}
The various functions of temperature involved in the expressions above are 
given in Appendix~A.

The $T$-dependence  and the evolution along the perturbation 
expansion of the mass and the "temporal" pion decay constant are displayed 
in Fig.3-4 in the case of the physical pion mass and pion decay constant: 
$M_\pi \simeq 140$ MeV and $F_\pi \simeq 93$ MeV. For both $M_\pi(T)$ and 
$F_\pi^t(T)$  the third order corrections have the opposite sign as those of the 
second order. 
At one loop, the mass is enhanced by the effects of the temperature, but the 
two-loop corrections bring it down (Fig.3). This may be a reflection of the fact 
that at $T=0$, the first correction is negative for $M$ (\ref{eq:MF}). Exactly 
the opposite happens to the pion decay constant (Fig.4 and (\ref{eq:MF})). The 
difference between the "temporal" and "spatial" pion decay constants is always 
positive (cf Fig.~7 in the next Section).

\setcounter{equation}{0}
\section{Massless quarks}

In our problem, the small $M_{\pi}$ and fixed $T$ case is equivalent to the 
limit $T \gg M_{\pi}$: only the ratio $M_\pi / T$ is relevant. In the chiral 
limit, that is when the quark masses tends to zero, $M_{\pi}$ tends to zero. The 
expressions of the pion decay constants and mass are much more readable ($F 
\simeq 88$ MeV is the pion decay constant in the chiral limit~\cite{GeL}):

\begin{eqnarray}
\left. \frac{M_{\pi}^2(T)}{M_{\pi}^2} \right|_{\hat{m}=0}&=&1+\frac{T^2}{24 
F^2}+\frac{T^4}{36 F^4} \left[  
\frac{19}{480}+K + \ln \frac{T}{\mu}
\right.  \label{eq:chMt}\\
&&\left. \hspace{2cm}  -\frac{192 \pi^2}{5} (L_1^r(\mu)+2 
L_2^r(\mu)) \right] 
+O(T^6) ,\nonumber 
\end{eqnarray}

\begin{eqnarray}
\left. \frac{{\rm Re} \Big( F_{\pi}^t(T) \Big)^2}{F_{\pi}^2} 
\right|_{\hat{m}=0}&=&1-\frac{T^2}{6 F^2}+\frac{T^4}{36 F^4} \left[  
\frac{7}{60}-K - \ln \frac{T}{\mu}
\right.  \label{eq:chFtt}\\
&&\left. \hspace{2cm}  +\frac{192 \pi^2}{5} (L_1^r(\mu)+2 
L_2^r(\mu))\right] 
+O(T^6), 
\nonumber
\end{eqnarray}

\begin{eqnarray}
\left. \frac{ {\rm Re} \Big( F_{\pi}^t(T)-F_{\pi}^s(T) \Big)}{F_{\pi}} 
\right|_{\hat{m}=0}&=&\frac{T^4}{27 F^4} \left[  
-\frac{2}{15}-K - \ln \frac{T}{\mu}
\right.  \label{eq:chFst} \\
&&\left. \hspace{2cm}  +\frac{192 \pi^2}{5} (L_1^r(\mu)+2 
L_2^r(\mu))\right] +O(T^6). \nonumber 
\end{eqnarray}
Where the number
\begin{equation}
K=\ln2+\frac{1}{2} \Gamma'(1)+\frac{45}{\pi^4}\zeta'(4)-1.05\simeq-0.68
\end{equation}
contains the Euler gamma and Riemann zeta 
functions and a contribution from the integrals that had to be numerically 
evaluated. 
$\mu$ is the regularization scale used in~(\ref{eq:lamb}) and 
$L_i^r(\mu)=\bar{L}_i+\gamma_i  \; \ln \frac{M}{\mu} / (4 \pi)^2$ are the 
scale-dependent renormalized effective coupling constants. The scale dependence 
and the 
logarithmic divergences of the individual terms appearing 
in~(\ref{eq:chMt}-\ref{eq:chFst}) cancel, as they have to. 

Our expressions for the pion mass and decay constants can always be written in 
the form
\begin{equation}
\left. \frac{M_{\pi}^2(T)}{M_{\pi}^2} \right|_{\hat{m}=0} = 1+ 
\frac{T^2}{24 \; F^2}- \frac{T^4}{36 \; F^4}\ln \frac{\Lambda_M}{T}+O(T^6),
\end{equation}

\begin{equation}
\left. \frac{{\rm Re} \Big( F_{\pi}^t(T) \Big)^2}{F_{\pi}^2} 
\right|_{\hat{m}=0} = 1-\frac{T^2}{6 \; F^2} +\frac{ T^4 }{36 \; F^4} \ln 
\frac{\Lambda_t}{T}+O(T^6),
\end{equation}

\begin{equation}
\left. \frac{ {\rm Re} \Big( F_{\pi}^t(T)-F_{\pi}^s(T) \Big)}{F_{\pi}} 
\right|_{\hat{m}=0} = \frac{ T^4}{27 \; F^4} \ln 
\frac{\Lambda_{\Delta}}{T}+O(T^6).  \label{eq:lnfts}
\end{equation}
Where $\Lambda_{M,t,\Delta}$ are various scales which sizes are determined by 
the numbers and the values of the coupling constants appearing in 
(\ref{eq:chMt}-\ref{eq:chFst}). We take  the recent two-loop 
evaluation~\cite{Wan} as reference:
$\bar{l}_1=-1.7$ and $\bar{l}_2=5.4$, i.e. $32 \pi^2 (\bar{L}_1+2 \bar{L}_2) 
\simeq 1.66$. The scale we find are rather big compared to the ones usually 
involved in ChPT~\cite{GeL}: 
\begin{eqnarray}
\Lambda_M & \simeq & 1.9 \; {\rm GeV} \\
\Lambda_t & \simeq & 2.3 \; {\rm GeV} \\
\Lambda_{\Delta} & \simeq & 1.8 \; {\rm GeV}.
\end{eqnarray}

In the range allowed for the temperature, $\ln \Lambda_{M,t,\Delta}/T$ is 
positive. As already seen in the physical pion mass case, the third order 
corrections have the opposite signs to the second ones in $M_{\pi}(T)$ and 
$F_{\pi}^t(T)$, whereas  the latter is bigger than $F_{\pi}^s(T)$. This is shown 
in the Fig.5-7 both in the chiral limit and in the physical case. 

To see what becomes the GOR relation at order $T^4$, we need the quark 
condensate to $O(p^6)$. It has been computed to $O(p^8)$ in~\cite{GeL}. In the 
general case it reads:
\begin{eqnarray}
\hat{m} \langle \bar{q} q \rangle_T&=&\hat{m} \langle 0 | \bar{q} q |0 \rangle 
+\frac{3}{2} M_{\pi}^2 {\cal N}_{M_{\pi}}(T) \nonumber \\
&& +\frac{M_{\pi}^2}{F_{\pi}^2} \left[ M_{\pi}^2 {\cal N}_{M_{\pi}}(T) \left( 
-24 \bar{L}_4 +48 \bar{L}_6+48 \bar{L}_8+\frac{3}{64 \pi^2} \right) \right.  
\label{eq:qq} \\
&& \left. \hspace{1.5cm} +\frac{3 M_{\pi}^2}{4} {\cal N}_{M_{\pi}}(T) \; 
\partial_{M_{\pi}^2} {\cal N}_{M_{\pi}}(T)+\frac{3}{8} {\cal N}_{M_{\pi}}^2(T) 
\right] + O(p^8), \nonumber 
\end{eqnarray}
thus in the chiral limit
\begin{equation}
 \langle \bar{q} q \rangle_T \Big|_{\hat{m}=0}=\langle 0 | \bar{q} q |0 
\rangle \left\{ 1- \frac{T^2}{8 F^2}-\frac{T^4}{384 F^4} \right\} 
+O(T^6). 
\label{eq:chqq}
\end{equation}
Which means that a modified GOR relation can be written down for massless quarks 
at finite 
temperature:
\begin{equation}
\lim_{\hat{m} \rightarrow 0}\; \frac{ 
M_{\pi}^2(T) \; {\rm Re}[\left(F^t_{\pi}(T)\right)^2]}{\hat{m} \langle \bar{q} 
q \rangle_T}=-1+O(T^6) \label{eq:gorT}
\end{equation}

This was already noticed in~\cite{Pis} and~\cite{Dom} and is a 
consequence of the Goldstone theorem at finite temperature as will be seen in 
the next Section. 

As expected, the essential characteristics of the low temperature behaviour of 
the three quantities examined are already present when the quark masses are sent 
to zero. Moreover, in this approximation a nice interpretation of 
the difference between the two pion decay constants and the imaginary parts of 
the residues of the two-point functions can be given. The next Section will come 
to that point.

Now, the symmetry groups occuring in the $O(4)$ Linear Sigma Model (LSM) are the 
same as those of QCD with two massless flavours. Hence the effective field 
theories of these two systems are identical, only their effective coupling 
constants differs. This implies that the results obtained above are valid as 
they stand also for the $O(4)$ model. In particular, the temperature expression 
of the pion mass contains a specific logarithmic contribution at order $T^4$. As 
it is absent in the LSM calculation described in~\cite{Pis,Ito}, their result is 
not complete at order $T^4$. The coupling constants $L_1^r(\mu)$, $L_2^r(\mu)$ 
can be evaluated for the LSM with~\cite{GaL}. The logarithmic contributions may 
be viewed as arising from a temperature dependent effective coupling constant of 
the LSM. Indeed, using the formula (34) in~\cite{Pis} 
\begin{equation}
\frac{M^2_\pi(T)}{M^2_\pi}=1+\frac{T^2}{6 \; F^2}-\frac{3 \pi^2}{15} \; 
\frac{T^4}{F^2 \; m_\sigma^2}, \label{eq:mlsm}
\end{equation}
and replacing $m_\sigma$ with $m_\sigma(T)$ defined by
\begin{equation}
\frac{1}{m^2_\sigma(T)}:=\frac{1}{m^2_\sigma}-\frac{5}{24 \pi^2 \; F^2} \; \ln 
\frac{T}{\alpha \; m_\sigma}, \label{eq:msT}
\end{equation}  
one recovers our result, provided that $\alpha \simeq 0.68$. \\
The same reasoning may be applied to the expressions found in~\cite{Pis} for 
both $F_\pi^{s,t}(T)$. Again the logarithms occuring at order $T^4$ are missing. 
These expressions are compatible with ChPT if $m_\sigma$ is replaced by 
$m_\sigma(T)$ like in (\ref{eq:mlsm},\ref{eq:msT}) but with $\alpha \simeq 
0.62$. In the notation used above the difference between the two logarithmic 
scales arises from $\Lambda_M \neq \Lambda_t$. This difference also manifests 
itself in the temperature dependence of the quark condensate at order $T^4$. If 
the formulae in~\cite{Pis} were correct, the temperature expression of the quark 
condensate would not contain a term of order $T^4$, in contradiction with the 
old calculation described in~\cite{GeL}, where the result is given up to and 
including contributions of order $T^6 \ln T$.

\setcounter{equation}{0}
\section{Gell-Mann Oakes Renner relation at finite \boldmath{$T$}}

To see why the Gell-Mann Oakes Renner relation has to take the 
form (\ref{eq:gorT}) when $T\neq0$ and to compute the corrections in the quark 
mass is the main purpose
of this Section. A possible way to do this is to go back where it originates at 
$T=0$. It may be derived from two Ward Identities involving the 
quark condensate and both the axial and pseudoscalar two-point Green's 
functions. As already mentioned the difference 
between the $T=0$ and the $T\neq0$ cases is a change of manifold in the path 
integral formalism. Hence the derivation of Ward Identities from the generating 
functional of QCD at finite temperature go through the same steps as at $T=0$. 
In the final equalities the vacuum expectation values of the involved operators 
are just replaced by their thermal average. These Ward Identities will be used 
both as a consistency check of the whole computation and as a source of the 
generalized relation. They will have byproducts which are going to clarify the 
meaning of our results.

To construct the finite temperature sisters of the Ward Identities which lead 
to the GOR relation at $T=0$, one has to use the axial-vector 
current~(\ref{eq:ax}) and the pseudoscalar density
\begin{equation}
P^a(x)=\bar{q}(x) \frac{\tau^a}{2} i \gamma_5 q(x).
\end{equation}
With the QCD Lagrangian, the following Ward Identities are obtained:
\begin{eqnarray}
q^\mu \tilde{G}_{\mu\nu}(q,T)&=&2 \hat{m} \tilde{G}_\nu(q,T), \label{eq:WI1} \\
q^\mu \tilde{G}_{\mu}(q,T)&=& 2 \hat{m} \tilde{G}(q,T)+\frac{1}{2} \; \langle 
\bar{q} q \rangle_T \label{eq:WI2},
\end{eqnarray}
where $G_{\mu \nu}(x,T)$ is the axial two-point Green's function defined in 
Section~3,  \\ $ G_\mu(x-y,T) \; \delta^{a b}:=i  \langle T \; A_\mu^a(x) P^b(y) 
\rangle_T$ and $G(x-y,T) \; \delta^{a b}:=i  \langle T \; P^a(x) P^b(y) 
\rangle_T$.

Note that at $T=0$, both identities are fulfilled by ChPT. The first one implies 
that
\begin{equation}
2 \hat{m} G_\pi=M_\pi^2 F_\pi, \label{eq:gf}
\end{equation}
and together with (\ref{eq:WI2}) in the chiral limit, one finds the  GOR 
relation at zero temperature.

The new two-point functions are represented in the same way as the axial one 
(\ref{eq:Gmn}):
\begin{equation}
\tilde{G}_{\mu}(q,T)=\Phi^{AP}_{\mu \nu}(q,T)- \frac{f_\mu(q,T) \; g(q,T)}{q_0^2 
-\Omega^2(\vec{q},T)}, \label{eq:Gm} 
\end{equation}
\begin{equation}
\tilde{G}(q,T)=\Phi^{P}(q,T)- \frac{g^2(q,T)}{q_0^2 -\Omega^2(\vec{q},T)}. 
\label{eq:G} 
\end{equation}
$\Omega(\vec{q},T)$ is defined as in (\ref{eq:omega}) and the residue 
$g(\vec{q},T)$ 
similarly as the one of the axial two-point function (\ref{eq:residue}). Again 
the strength of the coupling of the pseudoscalar density to the pion is defined 
to be $G_\pi(T):=g(\vec{q},T) \Big|_{\vec{q}=0}$, like in the axial case.

The first Ward Identity implies that
\begin{equation}
\Omega^2(\vec{q},T) \; f_t (\vec{q},T)-\vec{q}^{\; 2} \; f_s (\vec{q},T)=2 
\hat{m}\; g(q,T). \label{eq:Omf}
\end{equation}
At $\vec{q}=0$, it generates a relation very similar to (\ref{eq:gf}):
\begin{equation}
2 \hat{m}  G_\pi(T)=\Omega^2(\vec{q},T) \Big|_{\vec{q}=0} F_\pi^t(T) 
\label{eq:GFT}.
\end{equation}

Then the second Ward Identity together with (\ref{eq:Omf}) implies that
\begin{equation}
-F_t(\vec{q},T) \; g(\vec{q},T)+q^\mu  \Phi^{AP}_\mu (q,T)=2 \hat{m} 
\Phi^P(q,T)+\frac{1}{2} \langle \bar{q} q \rangle_T.
\end{equation}
The quark condensate is independent of $q$. the previous equality can thus be 
evaluated in the chiral limit and at $q=0$ (or conversely). Together with 
(\ref{eq:Omf}) it gives a generalization of the GOR relation at finite 
temperature in a form very 
close to the one at $T=0$ (and to all orders in the perturbation theory):
\begin{equation}
\lim_{\hat{m} \rightarrow 0}\; \frac{ 
\Omega(\vec{q},T) \Big|_{\vec{q}=0} \; \left(F^t_{\pi}(T)\right)^2}{\hat{m} 
\langle \bar{q} q \rangle_T}=-1.
\end{equation}
Because the quark condensate is a real quantity, the following relation must 
hold:
\begin{equation}
\lim_{\hat{m} \rightarrow 0}\; \frac{ {\rm Im} \Big[ 
\Omega(\vec{q},T) \Big|_{\vec{q}=0} \; \left(F^t_{\pi}(T)\right)^2 
\Big]}{\hat{m}}=0.
\end{equation}
Our result explicitly verifies this property at $O(p^6)$.

In order to compute the first corrections in the quark mass to the GOR relation 
at finite temperature, it is enough to compute the pseudoscalar two-point 
function. It requires the same building blocks as those of 
Section~3. Therefore only the end result will be given here. The graphs involved 
are the same as those appearing in Figure~2. The vertices are in general 
different and the diagrams $(2.1,4.1,6.4,6.7,6.20)$ are zero, 
because ${\cal L}^{(2)}$ is only linear in the pseudoscalar external field.

To present the result a representation similar as the one we used for 
the axial case in Section~3 is used:
\begin{equation}
\tilde{G}(q,T)=-\frac{\alpha^P(q,T)}{q_0^2-\Xi(q,T)}+\rho^P(q,T) \label{eq:GP}
\end{equation}
The various functions appearing in the previous expression are found to be:
\begin{eqnarray}
\alpha^P (q,T)&=& G_{\pi}^2/B^2- {\cal N}_{M_{\pi}}(T)  \nonumber \\
& & \hspace{-1cm} + \frac{1}{F_{\pi}^2} \left[ M_{\pi}^2 \; \bar{L}_P \; 
{\cal 
N}_{M_{\pi}}(T)  - \frac{1}{2} M_{\pi}^2 {\cal N}_{M_{\pi}}(T) \; 
\partial_{M_{\pi}^2} 
{\cal 
N}_{M_{\pi}}(T) \right.  \\
&& \left.  \hspace{1cm} -\frac{5}{2}  {\cal N}_{M_{\pi}}^2(T)
+\frac{M_{\pi}^2}{3} \bar{I}(q,T) 
\right] +O(p^8), \nonumber \label{eq:ap}
\end{eqnarray}

\begin{equation}
\rho^P(q,T)=R^P(q)+\frac{1}{F_{\pi}^2} \left[ \bar{L}_{\rho} \; {\cal 
N}_{M_{\pi}} (T) -\frac{1}{6} \bar{I}(q,T) 
\right] +O(p^8), \label{eq:rp}
\end{equation}
where two new combinations of coupling constants have been used
\begin{equation}
\left\{
\begin{array}{lll}
\bar{L}_P&=&  -48 \bar{L}_1 -16 \bar{L}_2+104 \bar{L}_4-224 \bar{L}_6-112 
\bar{L}_8+\frac{41}{288 \pi^2} \\
\bar{L}_{\rho}&=&32 \bar{L}_6-8 \bar{L}_8 -\frac{3}{32 \pi^2}.
\end{array} \right.
\end{equation} 
The expression for $\Xi(q,T)$ is the same as (\ref{eq:xi}). This was a first 
test for the whole computation: the poles of the two Green's functions under 
consideration must be identical.

The corrections to the massless quark world can be computed. The first one that  
appears is
\begin{equation}
\frac{ \Omega(\vec{q},T) \Big|_{\vec{q}=0} \; 
\left(F^t_{\pi}(T)\right)^2}{\hat{m} \langle 
\bar{q} q \rangle_T}=-1+2 \frac{M_{\pi}}{F_{\pi}^4} \; \frac{\partial}{\partial 
q_0} \bar{I}_{00}(q,T) \Bigg|_{q_0=M,\vec{q}=0} +O(M_{\pi}^2 \ln 
\frac{M_\pi}{T},p^8), 
\label{eq:TGOR}
\end{equation}
the expansion of the last integral in terms of the pion mass is given in the 
Appendix~B. We get:
\begin{equation}
\frac{ {\rm Re} \Big[ \Omega(\vec{q},T) \Big|_{\vec{q}=0} \; 
\left(F^t_{\pi}(T)\right)^2 \Big]}{\hat{m} \langle 
\bar{q} q \rangle_T}=-1- \frac{M_{\pi} T^3}{F_{\pi}^4} \; \left( 
\frac{1}{24}-\frac{3}{4 \pi^4} \zeta(3) \right) +O(M_{\pi}^2,M_{\pi} T^3 
\ln \frac{M_{\pi}}{T}).
\end{equation}

A small deviation linear in the pion mass is the first that appears. It is 
of the order of $6 \%$ at $T=100$ MeV and around $20 \%$ at $150$ MeV. This is 
in contradiction to the QCD sum rule result obtained in~\cite{Dom}, where the 
first correction is quadratic in $M_{\pi}$.

The Ward Identities~(\ref{eq:WI1},\ref{eq:WI2}) can be used as a consistency 
check of the whole calcualtion. The expressions for the quark 
condensate~(\ref{eq:qq}), the axial~(\ref{eq:GA}) and the 
pseudoscalar~(\ref{eq:GP}) two-point Green's functions must fulfill the 
identities. It is here the case, irrespective of the specific forms of the 
different independent functions involved in them.  This is also the case if one 
includes the imaginary parts of the finite terms that were ignored (see 
Section~3). 

The relation (\ref{eq:Omf}) is very instructive in the chiral 
limit. It implies that
\begin{equation}
\Omega(\vec{q},T)  = 
\frac{\Theta^A_s(\vec{q},T)}{\Theta^A_t(\vec{q},T)} \; \vec{q}^{\;2}. 
\label{eq:speed}
\end{equation}
Thus the speed of the pions in the chiral limit is
\begin{equation}
v^2_\pi \Big|_{\hat{m}=0}=1-\frac{{\rm Re} \Big( F_\pi^t(T)-F_\pi^s(T) 
\Big)}{F_\pi} \Bigg|_{\hat{m}=0}+O(p^8).
\end{equation}
It has to be smaller than the speed of light. Hence the difference between the 
real part of the two pion decay constants has to 
be positive (this remark was already made in~\cite{Pis}). Our result verifies 
this property. But looking at Fig.~7 or at (\ref{eq:lnfts}), one sees that the 
square of the speed tends to zero at $T \simeq 160$ MeV and even turns negative 
beyond that point. This is of course not allowed and it defines a natural 
limitation of the $O(p^6)$ ChPT expansion.

Finally, because all the quantities involved in (\ref{eq:speed}) are complex, 
this identity is in fact a system of two equalities containing six unknowns: the 
real and imaginary parts of the pole and of the two residues. Hence in the 
chiral limit, ${\rm Im} F_\pi^{s,t}(T)$ are completely determined by the other 
four 
quantities. Three of them are given in this article, whereas the imaginary part 
of the pole has been deeply studied in~\cite{Sch}. As a consequence, the 
physical content of the imaginary part  of the residues encodes that of the 
other quantities involved, which physical interpretation is well known.

\setcounter{equation}{0}
\section{Summary and Conclusion}

The dynamic of a pion travelling through a gas of pions at low temperature (up 
to circa $150$ MeV) can be computed with the help of the ChPT Lagrangian. The 
knowledge of the effective mass is important to understand how this happens. The 
two different effective pion decay constants that appear because 
of the breaking of Lorentz symmetry by the equilibrium state are also 
quite meaningfull. They were derived here to a $T^4$ accuracy performing a 
two-loop calculation in ChPT. The temperature dependence of these interesting 
quantities is small. This is due to the size of the pion decay constant at 
$T=0$ which governs the ChPT expansion.

When the quark masses are sent to zero, contrary to the result obtained within 
the Linear Sigma Model in~\cite{Pis}, we find a logarithmic dependence in the 
temperature of the three mentionned observables at the two-loop order. 
The difference between the "temporal" and "spatial" pion decay constants is 
positive. This is related to the fact that the velocity of the pions in the gas 
is smaller than the speed of light. The Gell-Mann Oakes 
Renner relation is still satisfied, which is a reflection of the Goldstone 
theorem at finite $T$.

For realistic quark masses, the temperature dependence is quite obscure: it is 
contained in complicated integrals and some pictures are needed to see what 
happens. The behaviour of the pion decay constants and mass for the physical 
pion mass is not very different from the massless case. 

The first corrections to the GOR relation are linear in the pion mass, they have 
a small magnitude. This does not agree with the result given in~\cite{Dom} using 
QCD sum rules.

Finally the presence of massive particles in the thermal equilibrium state 
deserves a comment. Their effects have been carefully analysed in~\cite{GeL}. 
The lightest ones that appear in our case are $K(500)$ and $\eta(550)$. Their 
masses do not vanish in the chiral limit ($m_{u,d} \rightarrow 0$). In our range 
of temperature, they 
behave like a dilute gas. They are of course exponentially suppressed, but their 
effects become more and more important when the temperature is increased. In the 
order parameter they generate a contribution of $0.5 \%$ at $T=100$ MeV with 
respect to that of the pions, whereas at $T=150$ MeV it becomes of the order of 
$10 \%$. At $T \simeq 160$ MeV, the mean distance between the massive states  is 
approximately $1.6$ fm and the number of massive particles per unit volume is 
the same as the numbers of pions. This gives a limitation to the ChPT approach. 
In the present calculation this limitation shows up at a similar temperature:
the square of the velocity of propagation of the Goldstone bosons turns 
negative. Our results are meaningless beyond that point.

\setcounter{equation}{0}
\section*{Acknowledgements}

It is a pleasure to thank H. Leutwyler for many useful discussions and a 
critical reading of the manuscript, U. B\"urgi, J. Gasser and C. Hofmann for 
informative comments.

\appendix
\renewcommand{\theequation}{\Alph{section}.\arabic{equation}}
\setcounter{equation}{0}
\section{Properties of the \boldmath{$T$}-dependent integrals}

A direct computation of the closed-eye,the guimbard and the sunset graphs 
brings different complicated integrals. But taking into account the symmetry of 
the integrands one can reduces them to two independent integral forms.
A short notation is first introduced:
\begin{equation}
\langle f(q,k_1,k_2) \rangle := i \int \frac{d^dk_1 \; 
d^dk_2}{(2 
\pi)^{2d}} \; \tilde{\Delta}_{\beta} (k_1) \; \tilde{\Delta}_{\beta} (k_2) \; 
\tilde{\Delta}_{\beta} (q-k_1-k_2) \; f(q,k_1,k_2).
\end{equation}
Because $\langle f(q,k_1,k_2) \rangle =\langle f(q,k_2,k_1) \rangle= \langle 
f(q,q-k_1-k_2,k_2) \rangle$ and that $\langle k_1^2 \rangle=M^2 
I(q,T)-\Delta_{\beta}^2(0)$, the expressions for the closed-eye, the guimbard 
and the sunset can be simplified 
into~(\ref{eq:ces},\ref{eq:guis},\ref{eq:suns}).

These two integrals $I(q,T)$ and $I_{\mu \nu}(q,T)$ diverge in $d=4$ dimensions. 
They have to be regularized. They contain three types of contributions. The 
$T$-independent terms are not of interest here. The parts involving one Bose 
distribution diverge for $d=4$ and the ones with a product of two statistical 
weights are finite for $d=4$. 

The first function of interest can be written as:
\begin{equation}
I(q,T)=\bar{I}(q,T)-6 \lambda {\cal N}_M(T)-\frac{3}{16 \pi^2} \; {\cal 
N}_M(T), 
\end{equation}
where ${\cal N}_M(T)$, already introduced in (\ref{eq:nm}), is
\begin{equation}
{\cal N}_M(T)=\int \frac{d^4 k}{(2 \pi)^3} \delta (k^2-M^2) \; n_B(\omega_k) 
\label{eq:nmapp},
\end{equation}
and $\bar{I}(q,T)$ is a finite integral given by
\begin{eqnarray}
 \bar{I}(q,T)&=&3 \int \frac{d^4k}{(2 \pi)^3} \delta (k^2-M^2) n_B(\omega_k) 
\bar{J}((q+k)^2) \nonumber \\
&&+ 3  \int \frac{d^4k}{(2 \pi)^3} \delta 
(k^2-M^2) n_B(\omega_{k}) K(q+k,T) \nonumber \\
&&+  i \int \frac{d^4k_1 \; {\rm d}^4k_2}{(2 \pi)^6} \delta 
(k_1^2-M^2) n_B(\omega_{k_1}) \delta (k_2^2-M^2) n_B(\omega_{k_2})  
\label{eq:Ibar}\\
&&\hspace{4cm} \delta ((q-k_1-k_2)^2-M^2)  n_B(\omega_{q-k_1-k_2}), \nonumber 
\end{eqnarray}

The second function we are interested in is
\begin{eqnarray}
I_{\mu \nu}(q,T)&=&\bar{I}_{\mu \nu} (q,T) \nonumber \\
&& \hspace{-1cm} + \lambda \left( {\cal N}_M(T) 
\left[ g_{\mu \nu} (\frac{1}{3} q^2-\frac{5}{3} M^2) -\frac{4}{3} q_{\mu} 
q_{\nu} \right] -\frac{10}{3} \; N_{\mu \nu}(M,T) \right)   \\
&& \hspace{-1cm} +\frac{1}{16 \pi^2} \left( {\cal N}_M(T) \left[ \frac{1}{18} \; 
g_{\mu \nu} 
(q^2+M^2)-\frac{5}{9} \; q_{\mu} q_{\nu} \right]  -\frac{14}{9} \; N_{\mu 
\nu}(M,T) \right), \nonumber
\end{eqnarray}
where $N_{\mu \nu}(M,T)$ is
\begin{equation}
N_{\mu \nu} (M,T)= \int \frac{d^4k}{(2 \pi)^3} k_{\mu} 
k_{\nu} \delta (k^2-M^2)  n_B(\omega_k), \label{eq:nmn}
\end{equation}
and the finite integral $\bar{I}_{\mu \nu}(q,T)$ is
\begin{eqnarray}
\bar{I}_{\mu \nu}(q,T) &=& \int \frac{d^4k}{(2 \pi)^3} \delta (k^2-M^2) 
n_B(\omega_k) \bar{J}((q+k)^2) \nonumber \\
&&\hspace{2cm} \left\{ \frac{5}{3} k_{\mu} k_{\nu}+\frac{2}{3}q_{\mu} q_{\nu}  
+\frac{2}{3} (q_{\mu} k_{\nu}+q_{\nu} k_{\mu}) \right. \nonumber \\
&&\left.\hspace{2.2cm} + \left( \frac{M^2}{2} -\frac{q^2}{6} - \frac{qk}{3}  
\right) g_{\mu \nu} - \frac{2 M^2}{3} \; \frac{(q_{\mu}+k_{\mu}) 
(q_{\nu}+k_{\nu})}{(q+k)^2} 
\right\} \nonumber \\
&&+\int \frac{d^4k}{(2 \pi)^3} \delta 
(k^2-M^2) n_B(\omega_k)  \left\{ -2 k_{\mu} K_{\nu}(q+k,T) \right. \nonumber \\
&& \left. \hspace{2cm} +K(q+k,T) [4 k_{\mu} k_{\nu}+q_{\mu} q_{\nu} +2 
(q_{\mu} k_{\nu}+q_{\nu} k_{\mu})] 
   \right\} \label{eq:Ibarmn} \\
&&+  i \int \frac{d^4k_1 \; {\rm d}^4k_2}{(2 \pi)^6} \delta 
(k_1^2-M^2) n_B(\omega_{k_1}) \delta (k_2^2-M^2) n_B(\omega_{k_2}) \nonumber \\
&&\hspace{2cm} \delta ((q-k_1-k_2)^2-M^2)  n_B(\omega_{q-k_1-k_2})  \nonumber 
\\
&& \hspace{2.5cm} \left\{ 4 k_{1 \mu} k_{1 \nu}+q_{\mu} q_{\nu} -2 (q_{\mu} 
k_{1 \nu}+q_{\nu} k_{1 \mu}) +2 k_{1 \mu} k_{2 \nu} \right\}. \nonumber 
\end{eqnarray}

In (\ref{eq:Ibar},\ref{eq:Ibarmn}), three one-loop functions have been 
introduced to get a simpler representation. 
The usual $T=0$ one
\begin{equation}
\bar{J}(q^2)=-\frac{1}{16 \pi^2} \int_0^1 dx \ln(1-q^2 x 
(1-x)/M^2), \label{eq:jbar}
\end{equation}
together with temperature dependent ones:
\begin{eqnarray}
K(q,T)&=&i \int \frac{d^4 k}{(2 \pi)^3} \delta(k^2-M^2) \; n_B(\omega_k) \; 
\tilde{\Delta}(q-k) \nonumber \\
&=& \frac{1}{16 \pi^2 |\vec{q}|} \int_0^{\infty} dk \; \frac{k}{\omega_k} 
n_B(\omega_k) \ln \frac{(q^2-2 k |\vec{q}|)^2-4 \omega^2_k \; q_0^2}{(q^2+2 k 
|\vec{q}|)^2-4 \omega^2_k \; q_0^2} \\
&&+\frac{i}{8 \pi |\vec{q}|} \int_0^{\infty} dk \; \frac{k}{\omega_k} 
n_B(\omega_k) 
\nonumber
\end{eqnarray}
and
\begin{eqnarray}
K_{\mu}(q,T)&=&i \int \frac{d^4k}{(2 \pi)^3} \delta(k^2-M^2) \; n_B( \omega_k) 
\; \tilde{\Delta}(q-k) k_{\mu} \nonumber \\
&=& \frac{1}{\vec{q}^{\; 2}} \left( n_{\mu} [\frac{q_0}{2} {\cal 
N}_M(T)+\frac{q_0 q^2}{2} K(q,T)-q^2 K_0(q,T)] \right. \\
&&\left. \hspace{1cm} -q_{\mu} [\frac{1}{2} {\cal N}_M(T)+\frac{q^2}{2} 
K(q,T)-q_0 K_0(q,T)] \right), \nonumber
\end{eqnarray}
where $n=(1,0,0,0)$. An integral representation for $K_0(q,T)$ can be given:
\begin{equation}
K_0(q,T)=\frac{1}{16 \pi^2 |\vec{q}|} \int_0^{\infty} dk \; k \; n_B(\omega_k) 
\; 
\ln \frac{q^4-4 (\omega_k \; q_0-k |\vec{q}|)^2}{q^4-4 (\omega_k \; q_0+k 
|\vec{q}|)^2}
\end{equation}

Note that because of the symmetry of the problem under spatial rotations:
\begin{equation}
N_{ik}(M,T)=\frac{1}{3} \; \delta_{ik} \Big( N_{00}(M,T) - M^2 {\cal N}_M(T) 
\Big) ,
\end{equation}
and
\begin{eqnarray}
\bar{I}_{0i}(q,T)&=&q_i \; \bar{\kappa}(q,T)  / 4 \\
\bar{I}_{ik}(q,T)&=&\delta_{ik} \; \bar{I}_1(q,T)+ q_i  q_k \; \bar{I}_2(q,T). 
\label{eq:rot}
\end{eqnarray}

The examined observables are defined on-shell at $\vec{q}=0$ and $q_0=M$. In 
that special case, a simplified representation can be given. Note that 
because $\omega_q \geq M$, the different sums of $T$-independent one-loop 
functions 
appearing in the integrands can be rewritten as:
\begin{eqnarray}
\bar{J}(2 M^2+2 M \omega_k)+\bar{J}(2 M^2-2 M \omega_k)&=&\frac{1}{8 \pi^2} \; 
\left( 2+\frac{\omega_k}{|\vec{k}|} \; \ln \sigma(k) \right), \nonumber \\
\bar{J}(2 M^2+2 M \omega_k)-\bar{J}(2 M^2-2 M \omega_k)&=&\frac{1}{8 \pi^2} \; 
\left( 2+\frac{M}{|\vec{k}|} \; \ln \sigma(k) \right),
\end{eqnarray}
where
\begin{equation}
\sigma(k):=\frac{\sqrt{\omega_k+M}-\sqrt{\omega_k-M}}{\sqrt{\omega_k+M}+ 
\sqrt{\omega_k-M}}.
\end{equation}

The following representations can be obtained:
\begin{eqnarray}
 {\rm Re} [ \bar{I}(q,T)] \Bigg|_{q_0=M,\vec{q}=0}&=&\frac{3}{32 \pi^4} 
\int_0^{\infty} dk 
\frac{k^2}{\omega_k} n_B(\omega_k) \left( 2+\frac{\omega_k}{k} \log \sigma(k)
 \right)   \label{eq:RIbar} \\
&& +\frac{1}{16 \pi^4} \int_0^{\infty} dk \; \int_0^1 d 
\alpha \; \frac{\alpha k^5}{\omega_{\alpha} \omega_k} \; n_B(\omega) \; 
n_B(\omega_{\alpha}) \; \ln \frac{1+\alpha}{1-\alpha} \nonumber
\end{eqnarray} 

\begin{eqnarray}
{\rm Re} [\bar{I}_{00}(q,T)] \Bigg|_{q_0=M,\vec{q}=0}&=&\frac{1}{96 \pi^4} 
\int_0^{\infty} dk 
\frac{k^2}{\omega_k} n_B(\omega_k) \Bigg\{ (5 k^2+7 M^2)  \label{eq:RIbarmn} \\
&& \hspace{.5cm} \left( 2+\frac{\omega_k}{k} \log 
\sigma(k) \right)+2 \omega_k M \left( 2-\frac{M}{k} \log 
\sigma(k) \right) \Bigg\} \nonumber \\
&& +\frac{1}{16 \pi^4} \int_0^{\infty} d k \int_0^1 d\alpha \; \alpha k^3 
n_B(\omega_k) \; n_B(\omega_{\alpha}) \; \ln \tau(\alpha,k). \nonumber
\end{eqnarray}
The expressions given above involve some new functions:
\begin{equation}
\omega_{\alpha}:=\sqrt{\alpha^2 k^2 + M^2},
\end{equation}
\begin{equation}
\tau(\alpha,k):=\frac{2 k^2 \alpha^2+ M^2 (1+\alpha^2)-2 \alpha \omega_k 
\omega_{\alpha}}{2 k^2 \alpha^2+ M^2 (1+\alpha^2)+2 \alpha \omega_k 
\omega_{\alpha}}.
\end{equation}
In general these integrals cannot be algebraically evaluated, even if the mass 
is zero.

Finally $\bar{I}_{ik}(q,T) \Big|_{q_0=M,\vec{q}=0}$ can be 
expressed in term of other known functions because of (\ref{eq:rot}):
\begin{equation}
\bar{I}_{ik}(q,T) \Big|_{q_0=M,\vec{q}=0}=\frac{1}{3} \; \delta_{ik} \; \Big( 
\bar{I}_{00}(q,T) -M^2 \bar{I}(q,T) + {\cal N}_M^2(T) \Big) 
\Big|_{q_0=M,\vec{q}=0} 
\end{equation}

\setcounter{equation}{0}
\section{Some functions in the chiral limit}

As already mentionned all our integrals are in fact functions of $M/T$. It is 
however not so easy to evaluate the desired ones in the chiral limit. Some 
clever tricks can be found in~\cite{GeL}. 
The value of the four 
functions~(\ref{eq:nm},\ref{eq:nmapp},\ref{eq:RIbar},\ref{eq:RIbarmn}) in the 
chiral 
limit $M \ll T$ is:
\begin{eqnarray}
 {\cal N}_{M}(T) \Big|_{\hat{m}=0}&=& \frac{T^2}{2 \pi^2} \; \int_0^{\infty} 
dt \; t \; n_B(t T)-\frac{M}{2 \pi^2 \; T} \; \int_1^{\infty} \; dt \; 
\frac{1}{t \; \sqrt{t^2-1}}  +O(\frac{M^2}{T^2} \ln \frac{M}{T})  \nonumber  \\
&=&\frac{T^2}{12}-\frac{M 
T}{4 \pi}+O(\frac{M^2}{T^2} \ln \frac{M}{T}), 
\end{eqnarray}

\begin{eqnarray}
 N_{00}(M,T) \Big|_{\hat{m}=0}&=& \frac{T^4}{2 \pi^2} \; \int_0^{\infty} dt 
\; t^3 \; n_B(t T) +O(\frac{M}{T})\nonumber \\
&=&\frac{\pi^2 T^4}{30}+O(\frac{M^2}{T^2}),
\end{eqnarray}

\begin{equation}
 {\rm Re} [\bar{I}(q,T)] \Bigg|_{q=0,\hat{m}=0}=\frac{T^2}{64 \pi^2} \;\ln 
\frac{M}{T} + O(\left(\frac{M}{T} \right)^0),
\end{equation}

\begin{eqnarray}
 {\rm Re} [\bar{I}_{00}(q,T)] \Bigg|_{q=0,\hat{m}=0}&=& \frac{1}{2} 
\ln \frac{M}{T} \; \left(  {\cal N}_{M}(T) \Big|_{\hat{m}=0} \right)^2 
\nonumber \\
&&+\frac{5}{48 \pi^2} (\ln \frac{M}{2 T}+2)  N_{00}(M,T) \Big|_{\hat{m}=0}
\nonumber \\
&&  +\frac{T^4}{2}\; \ln \frac{M}{2 T} \; \Big( {\cal N}_M (T) \Big)^2 \\
&&+\frac{T^4}{8 \pi^4} \int_0^{\infty} dt \; \int_0^1 d 
\alpha \; \alpha t^3 \; n_B(tT) \; n_B(\alpha t T) \; \ln 
\frac{1+\alpha^2}{\alpha^2 k^2} \nonumber \\
&=&T^4 \left( 
\frac{1}{144} \ln \frac{M}{T}-\frac{1}{144} \ln 
2+\frac{1}{1728} \nonumber \right. \\
&&\left. \hspace{1.5cm} -\frac{\Gamma'(1)}{288}-\frac{5 \zeta'(4)}{16 
\pi^4}+0.0073 
\right)+O(\frac{M}{T}). 
\nonumber
\end{eqnarray}

The numbers appearing in the previous expressions are due to numerically 
evaluated integrals that come directly from the representations shown in 
Appendix~A.

\begin{eqnarray}
 \bar{\kappa}(q,T) 
\Bigg|_{q=0,\hat{m}=0} 
&=& \frac{1}{6 M} \left( \left. {\cal N}_{M}(T) \right|_{\hat{m}=0} 
\right)^2-\frac{5}{144 M \pi^2} \left. N_{00}(M,T) \right|_{\hat{m}=0}\nonumber 
\\
&=& O(\left(\frac{M}{T}\right)^0).
\end{eqnarray}

\begin{eqnarray}
 \frac{\partial}{\partial q_0} {\rm Re}[\bar{I}_{00}(q,T)] 
\Bigg|_{q=0,\hat{m}=0}
&=& \frac{1}{2 M} \left( \left. {\cal N}_{M}(T) \right|_{\hat{m}=0} 
\right)^2-\frac{5}{48 M \pi^2} \left. N_{00}(M,T) \right|_{\hat{m}=0} \nonumber 
\\
&&\hspace{1cm} +\frac{3 T^3}{16 \pi^4} \int_0^{\infty} dt \; t^2 \; n_B(t T) 
\nonumber 
\\
&&\hspace{-2cm} =T^3 
\left (\frac{3}{8 \pi^4} \zeta(3) -\frac{1}{48 \pi} \right)
+O(\frac{M}{T} \ln\frac{M}{T}),
\end{eqnarray}
The cancellation of the $1/M$ terms in the last integral in the chiral 
limit is very important for the safe of the GOR relation at finite 
temperature and the $T^3$ term is responsible of the $M_{\pi} 
T^3$ corrections~(\ref{eq:TGOR}).

\section*{Figure Captions}

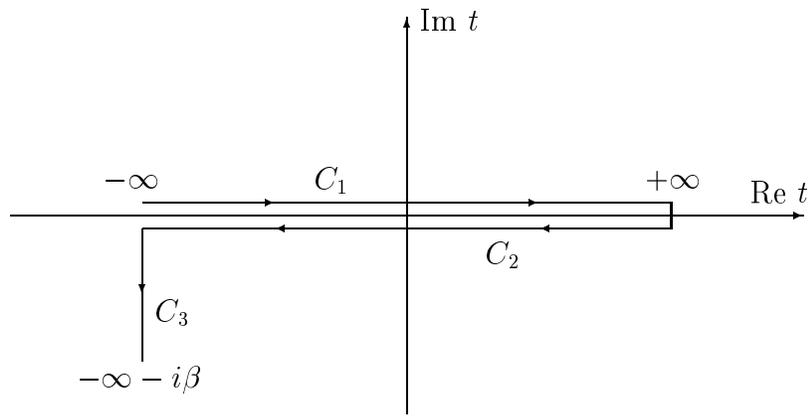
{\bf Fig.~1}: The Keldysh path of integration in the complex $t$ plane. The 
arrows indicate the time ordering on each part of the contour. Only $C_1$ and 
$C_2$ are relevant in the generating functional.\vspace{1cm} \\
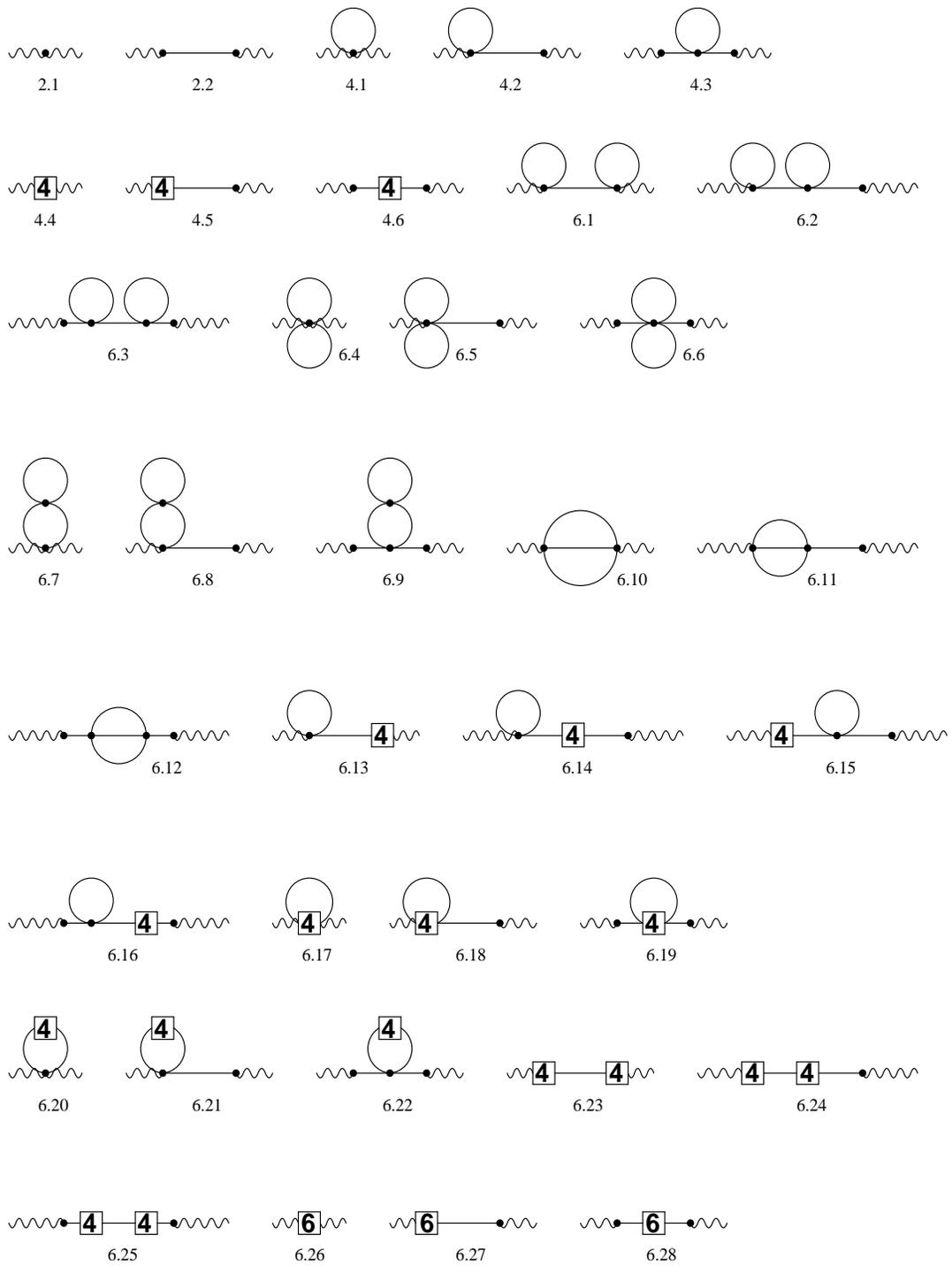
{\bf Fig.~2}: The Feynman diagrams necessary to compute the axial two-point 
Green's function to two loops. The various vertices correspond to the part of 
the effective Lagragian (\ref{eq:Leff}) involved: the "dot" for the ${\cal 
L}^{(2)}$ vertices, the "4" (resp. "6") for those of ${\cal L}^{(4)}$ (resp. 
${\cal L}^{(6)}$). The wiggled lines represent an external field, whereas the 
plain lines are the thermal propagators. The various parts of the RTF 
propagator and the crossed graphs are not explicitly given. 
\vspace{1cm} \\
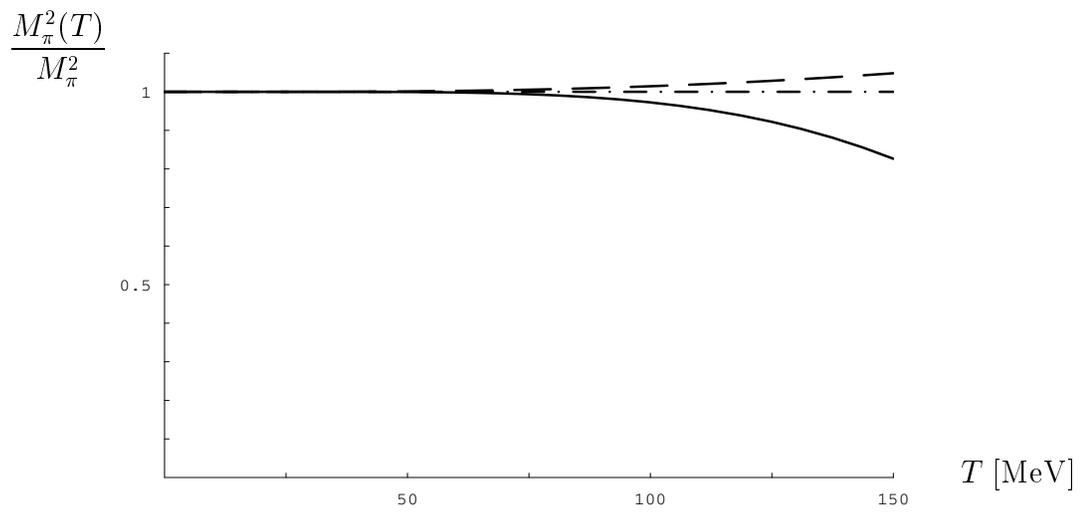
{\bf Fig.~3}: The effective pion mass at non-zero temperature. The dashed-doted 
curve represents the trivial result at the tree level, the dashed one the 
one-loop computation and the full one the two-loop approximation.\vspace{1cm} \\
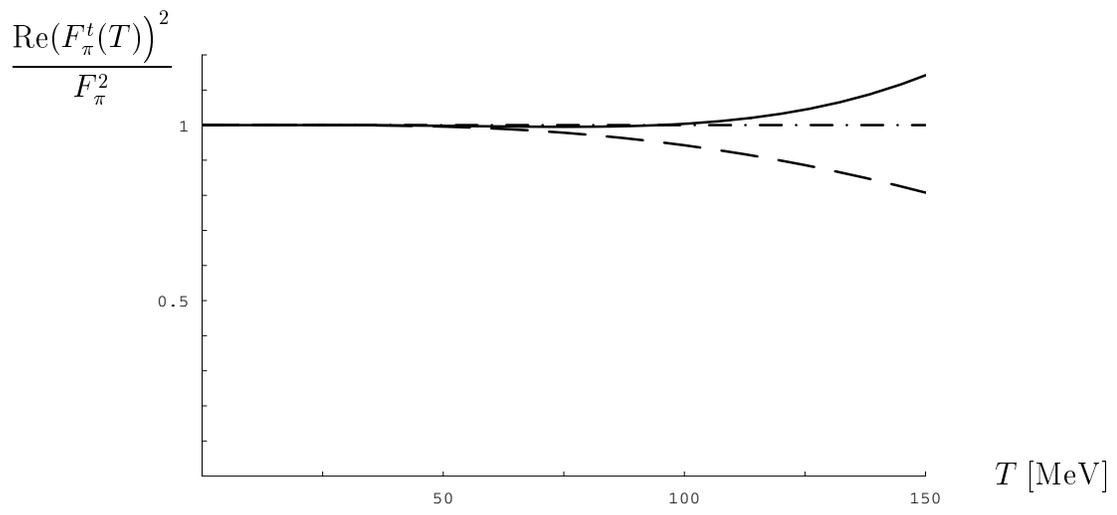
{\bf Fig.~4}: The real part of the "temporal"  pion decay constant at non-zero 
temperature. The dashed-doted curve represents the trivial result at the tree 
level, the dashed one the one-loop computation and the full one the two-loop 
approximation.\vspace{1cm}\\
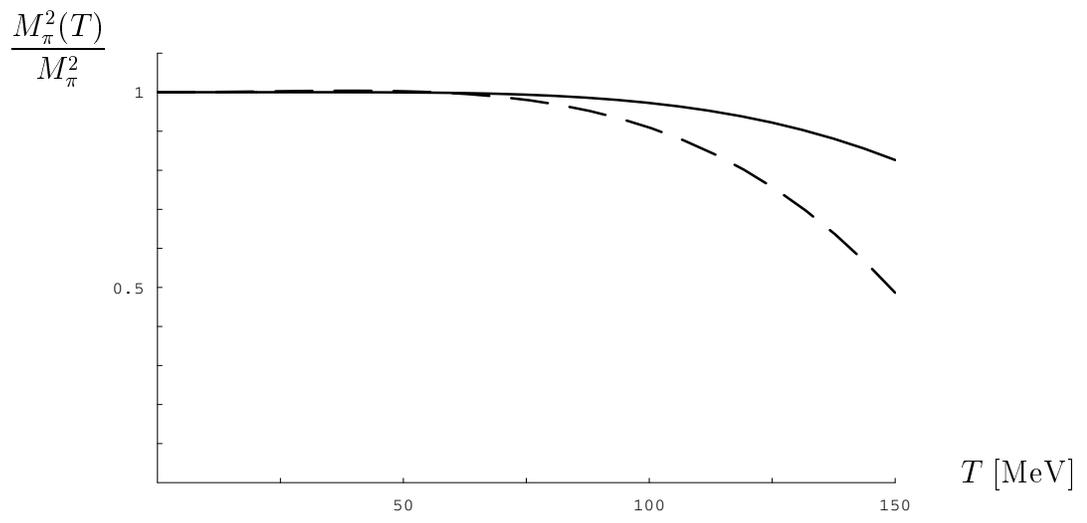
{\bf Fig.~5}: The $T$-dependent pion mass to two loops in the chiral limit 
(dashed curve) and in the physical case (full one).\vspace{1cm}\\
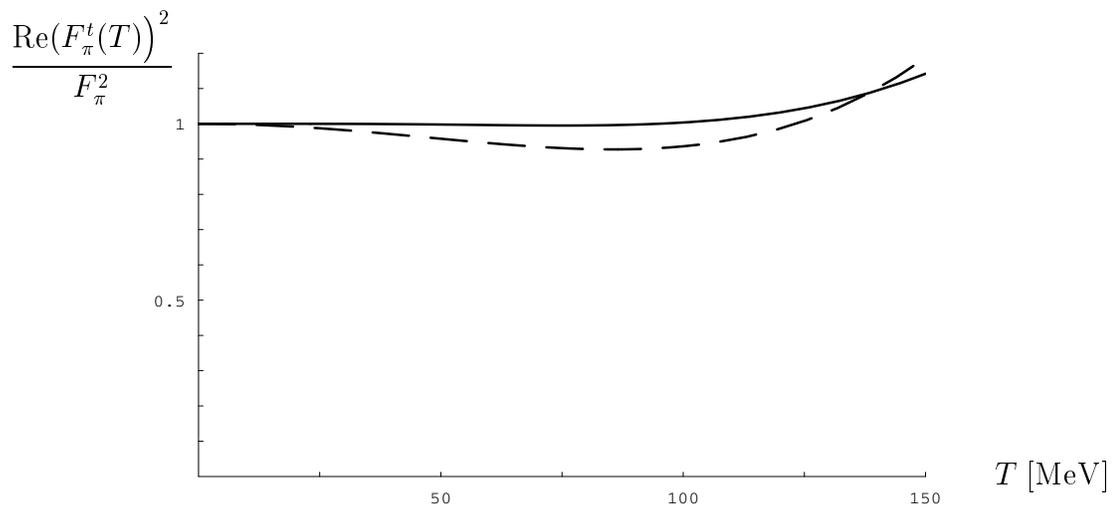
{\bf Fig.~6}: The real part of the "temporal" pion decay constant to two loops 
in 
the chiral limit (dashed curve) and in the physical case (full 
one).\vspace{1cm}\\
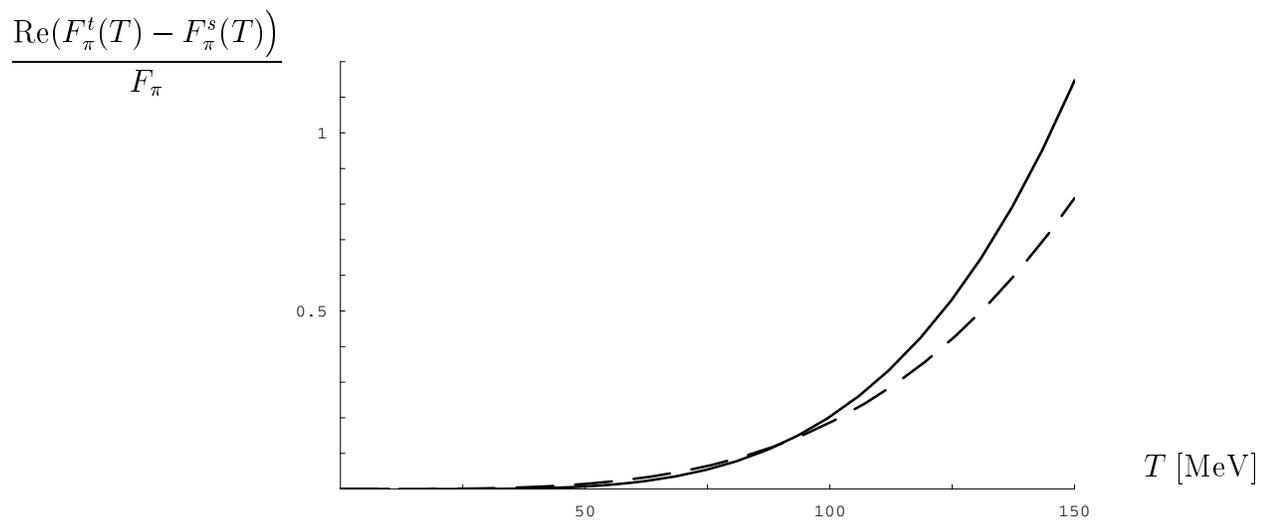
{\bf Fig.~7}: The difference between the real parts of the "temporal" and 
"spatial" 
pion decay constants to two loops in the chiral limit (dashed curve) and in the 
physical case (full one). In the chiral limit this quantity is related to the 
speed of the pions.


\begin{thebibliography}{99}
\bibitem{Smil} A.V. Smilga, "PHYSICS OF THERMAL QCD", TPI-MINN-96-23, 
hep-ph/9612347 and references therein.

\bibitem{GOR} M. Gell-Mann, R.J. Oakes and B. Renner, Phys. Rev. {\bf 175}, 2195 
(1968).

\bibitem{Wei} S. Weinberg, Physica A{\bf 96}, 327 (1979).

\bibitem{GaL} J. Gasser and H. Leutwyler, Ann. Phys. {\bf 158}, 142 (1984).

\bibitem{GaL3} J. Gasser and H. Leutwyler, Nuc. Phys. B{\bf 250}, 465 (1985).

\bibitem{GeL} P. Gerber and H. Leutwyler, Nuc. Phys. B{\bf 321}, 387 (1989).

\bibitem{Sch} A. Schenk, Phys.Rev. D {\bf 47}, 5138 (1993).

\bibitem{GaLT} J. Gasser and H. Leutwyler, Phys. Lett. B{\bf 184}, 83 (1987), 
and {\bf 188} (1987) 477.

\bibitem{Pis} R.D. Pisarski and M. Tytgat, Phys. Rev. D {\bf 54}, 2989 (1996).

\bibitem{LNR} H. Leutwyler, Phys. Rev. D{\bf49},3033 (1994). 

\bibitem{TFT} N.P. Landsman and Ch.G. van Weert, Phys. Rep. {\bf 145},141  
(1987) and references therein;\\
A.J. Niemi and G.W. Semenoff, Ann. Phys. {\bf 152}, 105 (1984); Nuc. Phys. B 
{\bf 230}, 181 (1984);\\
T. Altherr, Int. J. Mod. Phys. A{\bf 8}, 5605 (1993).

\bibitem{Bur} U. B\"urgi, Nuc. Phys. B {\bf 479}, 392 (1996).

\bibitem{L6} H.W. Fearing and S. Scherer, Phys. Rev. D{\bf 53},210 (1996).

\bibitem{Bij} J. Bijnens, G. Colangelo, G.Ecker, J. Gasser and M.E. Sainio, 
Phys. Lett. B{\bf 374}, 210 (1996).

\bibitem{Wan} G. Wanders, "CHIRAL TWO LOOP PION-PION SCATTERING PARAMETERS FROM 
CROSSING SYMMETRIC CONSTRAINTS", hep-ph/9705323.

\bibitem{Dom} C.A. Dominguez, M.S. Fetea and M. Loewe, Phys. Lett. B {\bf 387}, 
151 (1996).

\bibitem{Ito} H. Itoyama and A. H. Mueller, Nucl. Phys. B {\bf 218},349 (1984).

\end{thebibliography}
\end{document}